\documentclass[12pt,preprint]{aastex}
\usepackage{graphicx}
\usepackage{epsfig}
\usepackage{natbib}
\usepackage{amsmath, amsthm, amssymb}
\usepackage{longtable}
\usepackage{lipsum}
\usepackage{subfigure}

\title{The $\rm\gamma$-ray and Optical Variability Analysis of the BL Lac Object 3FGL J0449.4-4350}

\author{Xing Yang \altaffilmark{1}, Tingfeng Yi \altaffilmark{1,2}, Yan Zhang \altaffilmark{1}, Huaizhen Li \altaffilmark{3}, Lisheng Mao \altaffilmark{1}, Haiming Zhang \altaffilmark{4}, Li Ma \altaffilmark{3}}
\altaffiltext{1}{Key Laboratory of Colleges and Universities in Yunnan Province for High-energy Astrophysics, Department of Physics, Yunnan Normal University, Kunming 650500, China; yitingfeng98@163.com}
\altaffiltext{2}{Guangxi Key Laboratory for the Relativistic Astrophysics, Nanning 530004, China}
\altaffiltext{3}{Physics Department, Yuxi Normal University, Yuxi 653100, China}
\altaffiltext{4}{School of Astronomy and Space Science, Nanjing University, Nanjing 210023, China}
\begin{document}

\begin{abstract}
We have assembled the historical light curves of the BL Lac Object \object{3FGL J0449.4-4350} at optical and $\rm\gamma$-ray bands, the time spanning about 10 years, analyzed the periodic variability of the light curves by using four different methods (Lomb-Scargle periodogram, REDFIT38, Jurkevich and DACF). We detected a marginally possible quasi-periodic oscillation (QPO) of $\sim450$ days. Assuming it originates from the helical motion jet in a supermassive binary black hole (SMBBH) system undergoing major merger, we estimate the primary black hole mass $M\sim 7.7 \times 10^{9} \ \rm{M_{\bigodot}}$.
To explore the origin of the $\gamma$-ray, we investigated the optical-$\gamma$-ray correlations using discrete correlation function (DCF) method, and found that the correlation between the two bands is very significant. This strong correlation tends to imply lepton self-synchro-Compton (LSSC) model to produce the $\gamma$-ray.

%In the end, the possible explanations for the quasi-periodic timescale were discussed.
%The confidence of this periodicity is $99.7\%$ ($\sim3\sigma$) evaluated by carrying out Monte Carlo simulation.

\end{abstract}


\section{INTRODUCTION}
Blazars are one type of active galactic nuclei (AGNs) with a relativistic jet toward
the observers (e.g., Urry \& Padovani 1995).
According to the features of emission lines, blazars are divided into two groups: BL Lacertae
objects (BL Lacs; having weak or no emission lines) and flat-spectrum radio quasars (FSRQs; having strong emission lines).
FSRQs are usually the low synchrotron-peaked blazars (i.e., the synchrotron peak frequency
$\nu_{syn} <10^{14}$ Hz). BL Lacs are divided into three subclass: LBLs (low synchrotron-peaked BL Lacs, LBLs, $\nu_{syn} <10^{14}$ Hz), IBLs (intermediate peak frequency, $10^{14}<\nu_{syn} <10^{15}$ Hz)), and HBLs (high peak frequency, $\nu_{syn}  >10^{15}$ Hz) (Abdo et al. 2010).

The variabilities are important features of Fermi blazars. The observed multi-wavelength variabilities of many sources show
some kind of periodicity (Valtaoja et al. 1985; Sillanpaa et al. 1996; Li et al. 2006; Xie et al. 2008; Zhang et al. 2009; Sandrinelli et al. 2016A, B; Zhang et al. 2017A, B, C), e.g. PKS 2155-403 (Zhang et al. 2017A; Sandrinelli et al. 2014, 2016A), PG 1553+113 (Ackermann et al. 2015A; Sobacchi et al. 2017; Yan et al. 2018; Zhou et al. 2018). However, many of these claims are marginal, Most of them cannot be verified by new observations. Although various processes can lead to different quasi-periodic oscillation (QPO), the most of temporal variability of blazars is
essentially stochastic. The QPOs in blazars are rare and transient in nature (Gupta et al. 2019). The mechanism leading to the temporal variability remains completely unsettled.

3FGL J0449.4-4350 (PKS 0447-439) is a TeV BL Lac (HBL, $\rm Log \nu_{syn}  =15.671$ (Hz))(H.E.S.S. Collaboration 2013), with redshift $z = 0.205$ and has a spectral index $\Gamma=1.85$, in the third Fermi-LAT AGN catalog (3LAC; Ackermann et al. 2015B).
To search quasi-periodicity, we assembled the historical light curve data of \object{3FGL J0449.4-4350} at
optical from Catalina Real-time Transient Survey (CRTS) and $\rm\gamma$-ray bands (with the time spanning about 10 years) from Fermi LAT data center.
This source drastically varies in optical band and $\gamma$-ray, with the similar duration of the flares.
A interesting phenomenon of the 3FGL J0449.4-4350 light curves is that the optical band and $\gamma$-ray flux synchronously change,
which shows that long-term monitoring is essential for understanding the typical behaviour in this source.

In this paper, we investigate the time series analysis of these light curves in search of quasi-periodicity using several different methods, and detect a possible QPO of about 450 days. In Section 2, we briefly describe the $\gamma$-ray Fermi LAT data and our analysis procedure. In Section 3, we present the QPO search methods we employed and the results of those analyses. In Section 4, we present detailed correlation analysis of the Fermi-LAT $\gamma$-ray and optical V-band light curves. A summary and discussion are given in Section 5.

\begin{figure}
\centering
\includegraphics[angle=0,scale=0.8]{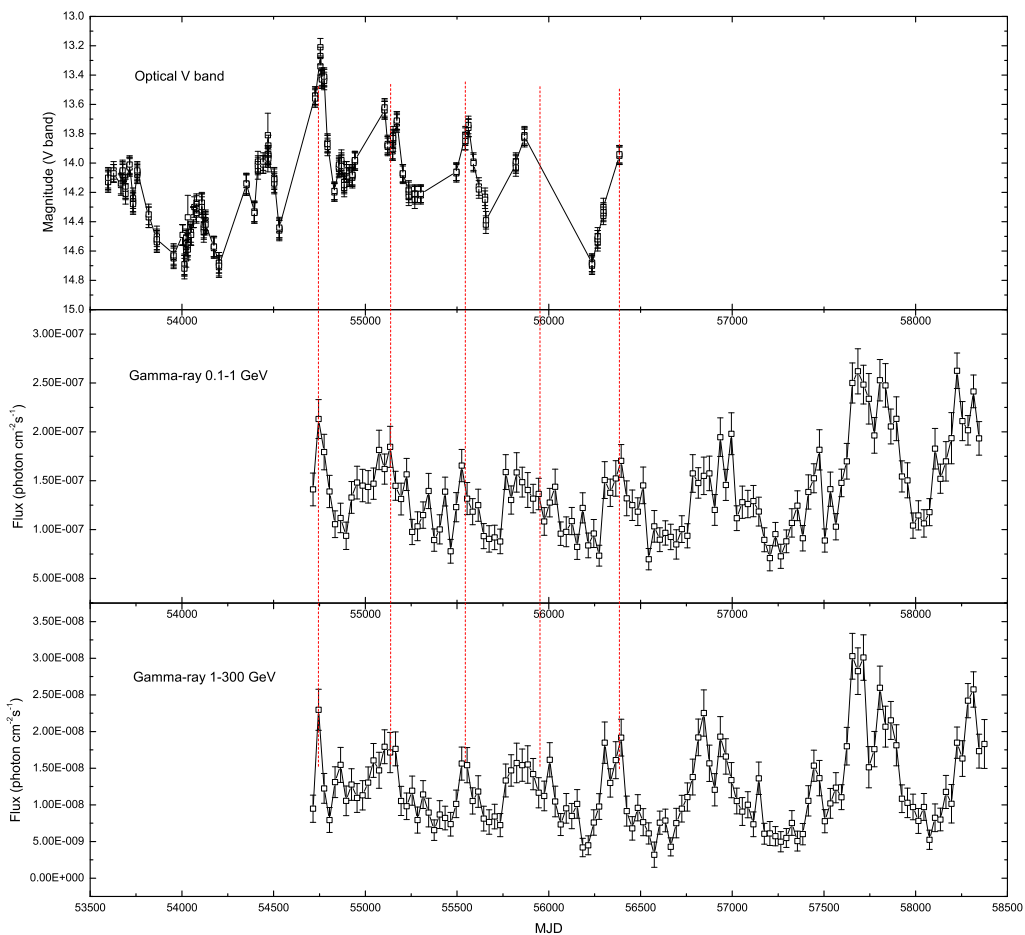}
\caption{The light curves of optical V, 0.1-1.0 GeV, 1-300 GeV band in 30 days bin for 3FGL J0449.4-4350. The vertical dash lines show clearly the similarity of the variations in three light curves.}
\end{figure}

\begin{figure}
\centering
\includegraphics[angle=0,scale=0.36]{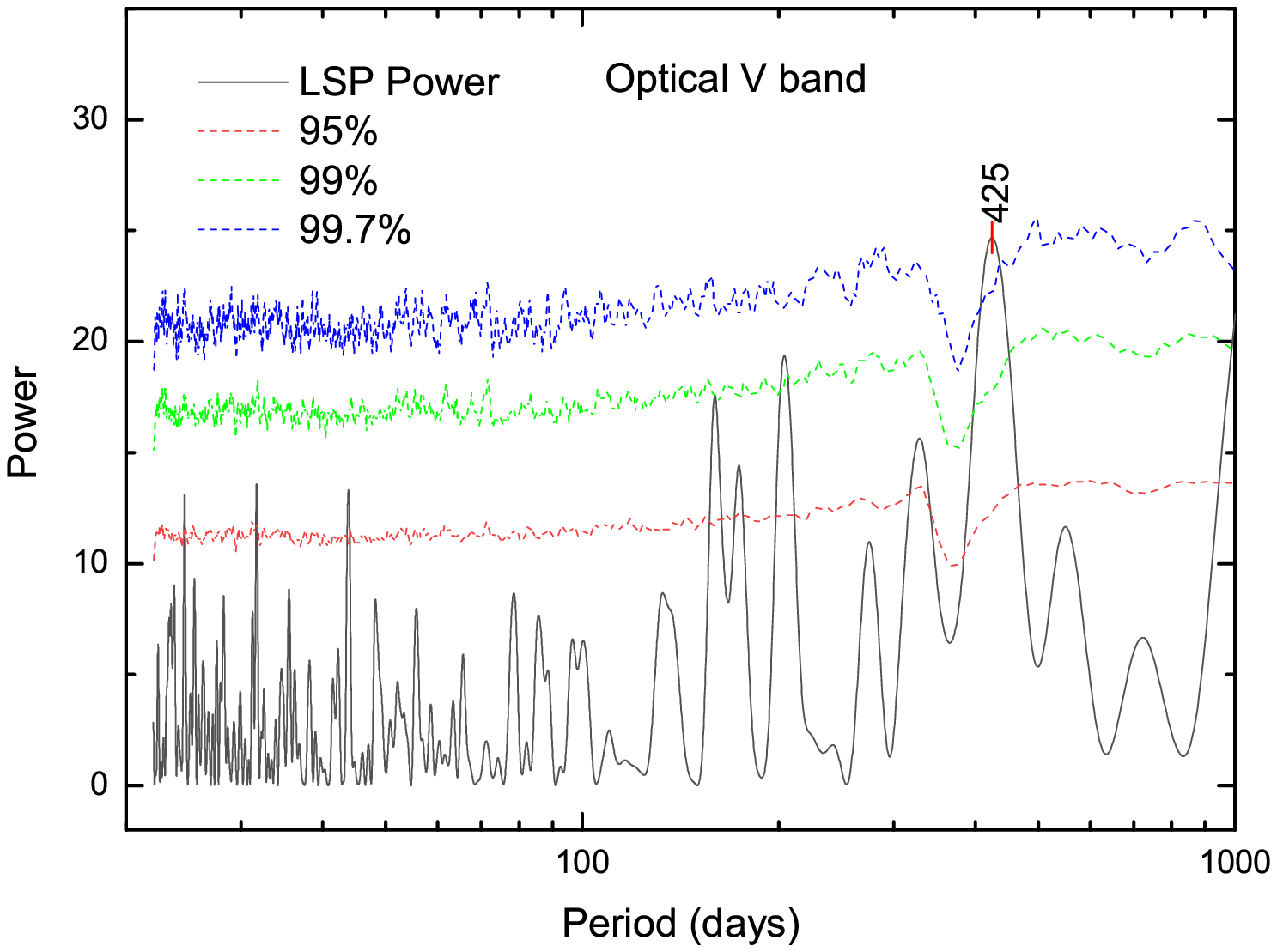}
\includegraphics[angle=0,scale=0.33]{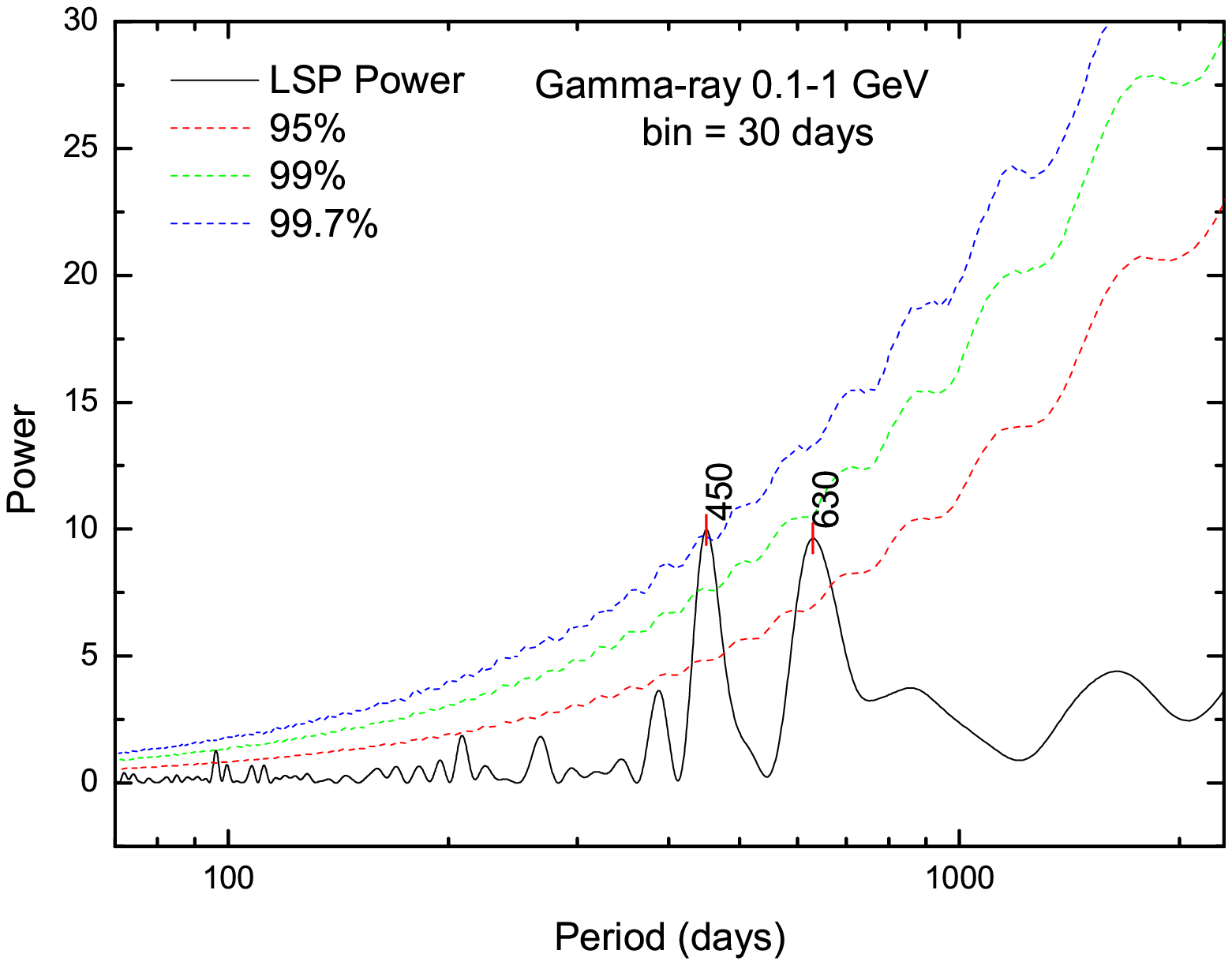}
\includegraphics[angle=0,scale=0.33]{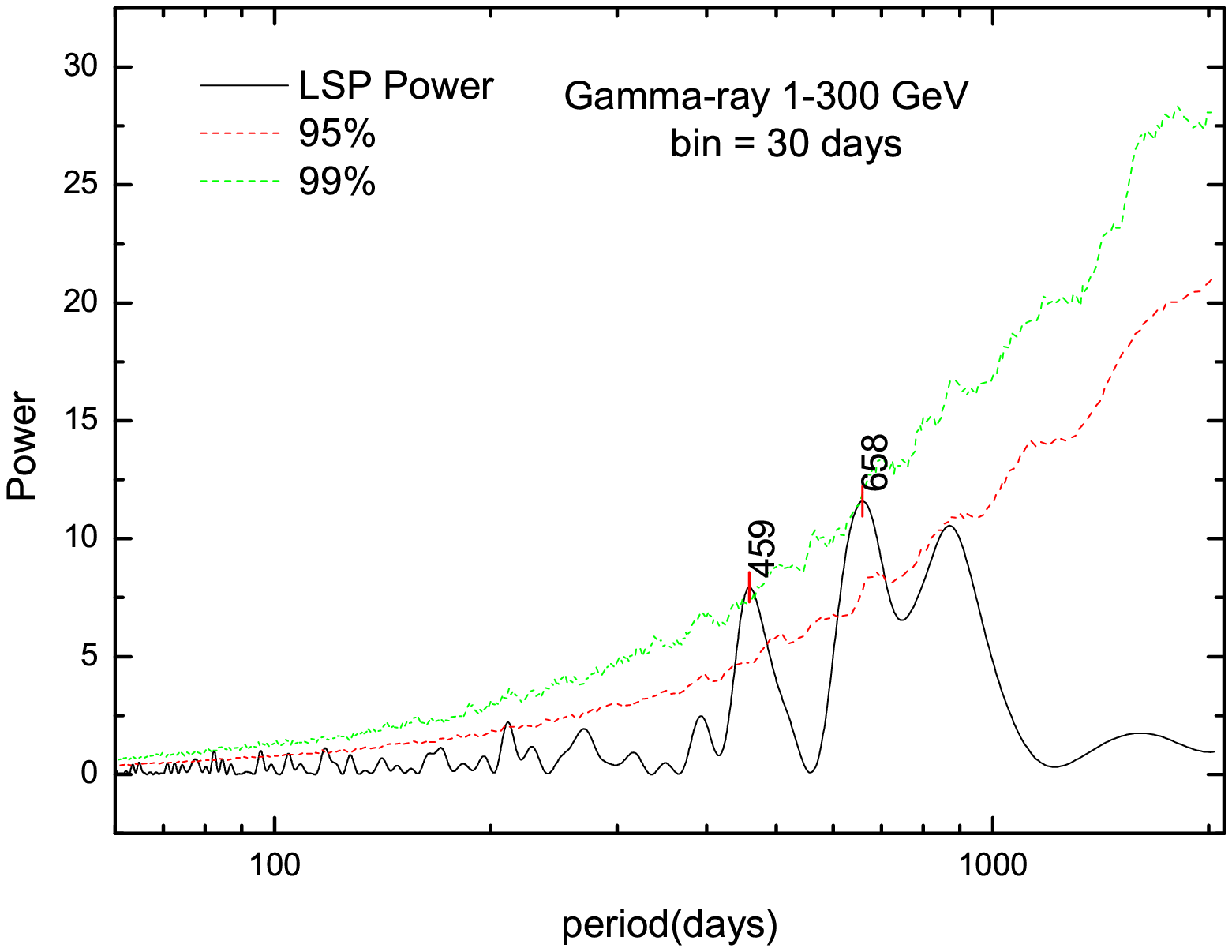}

\caption{The LSP power of $\rm\gamma$-Ray band (1.0-300 GeV) with 30 days bin and optical V-band light curves for 3FGL J0449.4-4350. The red, green, and blue dashed lines represent the confidence level of $95\%$, $99\%$  and $99.7\%$ respectively.}
\end{figure}

\section{OBSERVED DATA}
We collected the historical light curves data of $\rm\gamma$-ray and optical band for 3FGL J0449.4-4350.
The light curves are shown together in Fig. 1.
The data sets of each light curve are described in the following section respectively.
\subsection{OPTICAL DATA}
The provenance of the optical data set and characters are describe as follow.
The optical V band data is taken from the The Catalina Real-Time Transient Survey (CRTS)(Object-ID: SSS\_J044924.7-435008)\footnote[1]{http://nesssi.cacr.caltech.edu/DataRelease/}(Drake et al. 2009). The number of data points is 253.
The light curve of V band is also shown in Fig. 1 and the periodicity analysis is shown in Figure 2. For the calculation of the discrete correlation function (DCF) between the fluxes of two bands, we converted V band data from magnitudes to absolute flux densities using 3640 Jy as absolute flux density for zero-point
(M$_{\rm V}$ = 0) (Bessell 1979).

\subsection{FERMI/LAT DATA REDUCTION}

The $\gamma$-ray data of 3FGL 0449.4-4350 was downloaded from the Large Area Telescope (LAT) (Atwood et al. 2009, Abdo et al. 2009) on board Fermi satellite. We use the Fermi science tools software package (v11r06p03) to get the fluxes. Our light curves are produced in two different ways with the Fermi-specific tools gtlike/pyLikelihood that use likelihood analysis and gtbin employ the aperture photometry, respectively. We run the science tools gtlike/pyLikelihood to calculate the influence of the source. The model file is composed of all known 3FGL sources derived from 8 years of survey data in the ROI. The P8R2-SOURCE-V6 set of instrument response functions (IRFs) was used. The corresponding good time intervals (GTIs) were generated using the flag "(DATA $\rm QUAL>0$) \&\& (LAT $\rm CONFIG==1$)".
We select the photon energies between 100 MeV to 1 GeV and 1 GeV to 300GeV from MJD 54690 (2008-08-12) to MJD 58495 (2019-1-12) with 30 days bin. The region of interest (ROI) is $10^{\circ}$ radius with the center of source location (RA: 72.3529, DEC: $-43.8358$), a maximum zenith angle restriction of $105^{\circ}$ was applied to avoid the influence of $\rm\gamma$-rays from the Earth's limb. The $\gamma$-ray signal from the source is evaluated by gtlike/pyLikelihood with the maximum-likelihood test statistic (TS), We adopt the events with $\rm TS>21$.
The 30 days time bin aperture photometry light curves of  0.1-1 GeV, 1-300 GeV, are shown in Fig. 1. And the 7 days time bin gtlike light curves are shown in upper panel of Fig. 3. There is obvious variability in upper panel of Fig. 3, this pulsed variability is available for the periodicity analysis.

\section{SEARCH FOR PERIODICITY IN THE LIGHT CURVES}
To search for the possible QPOs in the light curves of 3FGL 0449.4-4350, Lomb-Scargle periodogram (LSP; Lomb 1976; Scargle 1982; Press et al. 1992) was performed.
For a time series $x(t_j)$ , $j=1,2, ... , N_0$, the LSP (Spectral power as a function of angular frequency $\omega$) is defined as:
\begin{equation}
P_{LS}(\omega)= \frac{1}{2}\left\{\frac{\sum_{j=1}^{N_{0}}(x(t_j)-\overline{x})\cos\omega(t_j-\tau)}{\sum_{j=1}^{N_{0}}\cos^{2}\omega(t_j-\tau)}+\frac{\sum_{j=1}^{N_{0}}(x(t_j)-\overline{x})\sin\omega(t_j-\tau)}{\sum_{j=1}^{N_{0}}\sin^{2}\omega(t_j-\tau)}\right\} ,
\end{equation}
where $\bar{x}\equiv\frac{1}{N_0}\sum_{j=1}^{N_{0}}(x(t_j)$. Here $\tau$ is defined by the relation
\begin{equation}
 \rm{tan}(2\omega\tau)=\frac{\sum_{j=1}^{N_{0}}\sin^{2}2\omega t_j}{\sum_{j=1}^{N_{0}}\cos^{2}2\omega t_j} .
\end{equation}

The light curves are commonly unevenly spaced in time, making it difficult to obtain an accurate estimate of their real spectrum according to the QPOs and to estimate the confidence level of the height of a peak in the power spectrum.
$P_{N}(\omega)=P_{LS}(\omega)/\sigma^{2}$, $\sigma^{2}\equiv\frac{1}{N-1}\sum_{j=1}^{N_{0}}(x(t_{i})-\overline{x})$, has an exponential probability distribution with unit mean (Horne \& Baliunas 1986). In other words, the probability that $P_{N}(\omega)$ will be between some positive $z$ and $z+dz$ is $e^{-z}$. If we scan some $N_{i}$ independent frequencies, the probability that none give values larger than $z$ is $(1-e^{-z})^{N_{i}}$, So
\begin{equation}
p(> z) \equiv 1-(1-e^{-z})^{N_{i}},
\end{equation}
is the false-alarm probability (FAP) of the null hypothesis (the data value are pure Gaussian noise), that is, the confidence level of any peak in $P_{N}(\omega)$.
At present, the whole topic of the independent frequency estimation is complex and not fully solved. In the literature, there are some methods proposed to try to deal with this problem (Horne \& Baliunas 1986; Pelt 1997; Jetsu \& Pelt 1999;  Reegen 2007; Zechmeister \& K\"{u}rster 2009; Vio et al. 2010, 2013; Olspert et al. 2018; VanderPlas 2018).
In order to evaluate the confidence level of the QPOs, we need to compute the number of independent frequencies $N_{i}$ in the LSP.
The $N_{i}$ usually depends on the number of frequencies sampled, the number of data points $N_0$, and their detailed spacing.

\begin{deluxetable}{llccll}
\tabletypesize{\scriptsize}
%\centering
\tablecaption{Independent Frequencies in Simulated Data}
\tablewidth{380pt}
\tablehead{ Bands    & $N_0$&  $N_i$  &   $N_i$ & $N_{sim}$  & Type of \\
                     &      &   (ofac=1, hifac=1) &  (ofac=4, hifac=1)&   & Spacing  \\}
\startdata

Optical V-band	&	253	&	89.79	&	180.79	&	$10^4$	&	clumps	\\
0.1-1.0 GeV	(30 days bin)&	122	&	64.88	&	136.34	&	$10^4$	&	even	\\
1.0-300 GeV (30 days bin)&	123	&	65.13	&	138.68	&	$10^4$	&	even	\\
0.1-1.0 GeV (7 days bin)&	471	&	241.17	&	563.11	&	$10^4$	&	random	\\
1.0-300 GeV (7 days bin)&	470	&	235.84	&	563.62	&	$10^4$	&	random	\\

\enddata
%\tablecomments{The data of }
\end{deluxetable}

We assume the span of a time series $x(t_j)$ is $T$, and the Nyquist frequency is $f_c= N_0/2T$.
For determining the independent frequencies $N_i$, we need do Monte Carlo simulation experiment. The sampling range of frequency is $1/T$ to Nyquist frequency $f_c$. In the Fast Fourier Transform (FFT) method, higher independent frequencies almost be integer multiples of $1/T$ , but more finely sampling than interval $1/T$ for LSP method. So, the number of different frequencies calculated by LSP method is $N_p=\frac{N_0}{2}(ofac\times hifac)$, where $ofac$ and $hifac=f_{hi}/f_c$ are two parameter, respectively describing over-sampling and highest frequency (with $f_c$ as a unit) to be calculated. If $ofac=1$ and $hifac=1$, we get $N_p=N_0/2$. But these different frequencies are not completely independent of each other.
In the Monte Carlo simulation experiment, we simulated a large number of pseudo-Gaussian noise data sets, with different spacings in the time coordinate, according to the sampling interval of multi wavelength observation data. The LSP of each data sets was computed, and the highest peak then was chosen in each LSP. As described by Horne \& Baliunas (1986), the false alarm function ($1-(1-e^{-z})^{N_{i}}$) was fitted to the highest peak distribution with $N_i$ as the variable parameter. The results of Monte Carlo simulation experiment are shown in Table 1. For $ofac=1$ and $hifac=1$, the $N_i$ in LSP of various multi-wavelength is about $N_0/2$, except the optical V band. As shown in the upper panel of Fig. 1, the
points of optical V band are closely clumped into groups of about 2.8. We also take $ofac=4$ and $hifac=1$, and obtain $N_i\approx 1.1N_0$ for even spacing. With $hifac$ unchanged, the value of $N_i$ gradually stabilizes to the number of peaks in the periodogram as $ofac$ increases. So, the independent frequencies $N_i$ may be estimated as the number of peaks in the periodogram (Zechmeister \& K\"{u}rster 2009).
The above discussion is based on the null hypothesis of independent Gaussian random (white) noise. If the measurement data is a signal plus red noise, the number of independent frequency in its LSP is even more difficult to estimate.

%The numerical simulations indicate that the consequences of unevenly sampled data seem  to concern the number of independent frequencies in periodogram.%To accurately estimate the confidence of periodic signals

However, if the observation data is a signal plus white noise, it has been pointed out in the literature that the number of independent frequencies is not a critical parameter to test the confidence level of a peak in LSP (Vio et al. 2010). In particular, empirical arguments indicate that this number can be safely set to $N_0$ (Press et al. 1992).
So far, only the predicted optical outburst of blazar OJ 287 has been confirmed by observation (Sillanpaa et al. 1988, 1996; Gupta et al. 2017).
The existence of the periodicity in the optical light curve observed in PG 1302-102 is yet in debate (Vaughan et al. 2016).
The most of the claimed periodic variability of AGNs can not withstand later observation (Covino et al. 2019), because the periodic variability maybe due to a red noise process. The random variations in light curves of blazars can be described as red noise with an approximately power law shape: $P(f) \propto (1/f)^{\beta}$. And the spectrum of red noise increase smoothly in power density to low frequencies.
We assessed the confidence of our findings by modelling the multi-wavelength variability as red noise with a power law index $\beta$.
To estimate the confidence, 10 000 light curves were
simulated by the method described in Timmer \& Koenig (1995) for each of the power law index $\beta$ values.
For obtaining the $\beta$ index of each wavelength band, we fitted the spectrum of the periodogram with power law, as shown in the first line panels of Fig. 4. Once the 10 000 light curves were simulated by using even sampling interval, and their LSP was computed. Consequently, using the spectral distribution of the simulated light curves, local 95, 99 and 99.7 percent confidence contour lines were evaluated.
But for optical V band (the points of this light curve are too clustered, and the gap is too uneven), using the sampling interval of original curve of light variation, we simulated 10 000 new light curves to calculate the confidence, because the uneven sampling of the light curve should be properly accounted along with red noise (Bhatta 2019).
As shown in Fig. 2, there are two  peaks in the periodogram, which hint two possible QPOs with $T1=450\pm23$ and $T2=630\pm58$ days for 0.1-1.0 GeV $\gamma$-ray band in 30 days bin. And similar periods in the 1-300 GeV $\gamma$-ray band with low confidence levels. For optical V-band, there is one peak with about $99\%$ confidence, which possible QPO of $425\pm40$ days, $\approx T1$. The half-width at the half-maximum (HWHM) of the peak was taken as a measure for the uncertainty in the value of QPO. The confidence basing on red noise and FAP basing on white noise are all shown in the Table 2. However, analytical estimates for FAPs for Lomb-Scargle periodogram are unreliable (VanderPlas 2018). The periodicity our discuss is just, at best, a hint of a detection.
%Since the $T1=450\pm23$ days is the most prominent period of $\gamma$-ray variabilities for 3FGL J0449.4-4350, we consider that $T1=450\pm23$ and $T2=630\pm58$ days are two major variability timescales in 3FGL J0449.4-4350

\begin{figure}
\centering
\includegraphics[angle=0,scale=0.48]{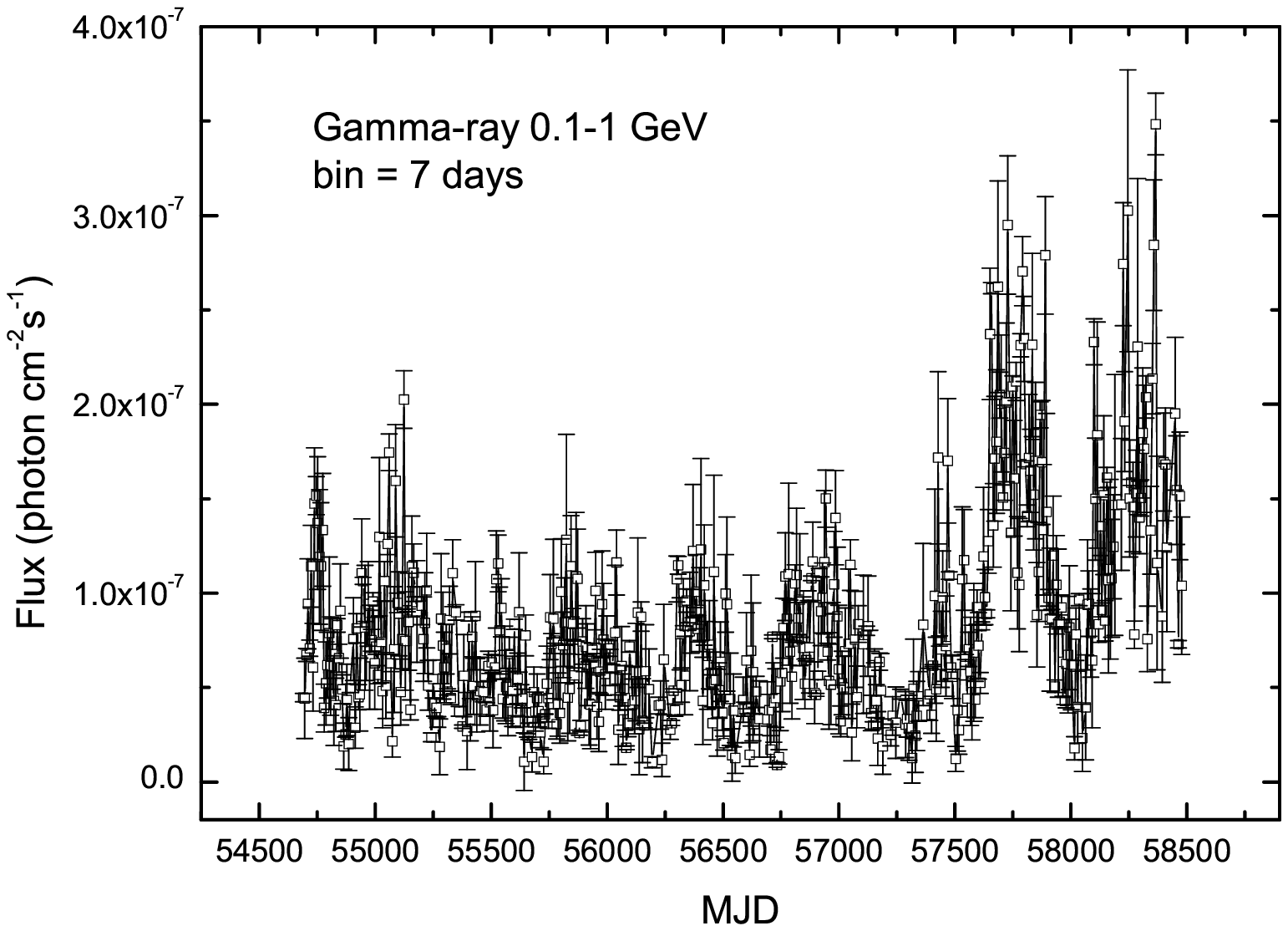}
\includegraphics[angle=0,scale=0.48]{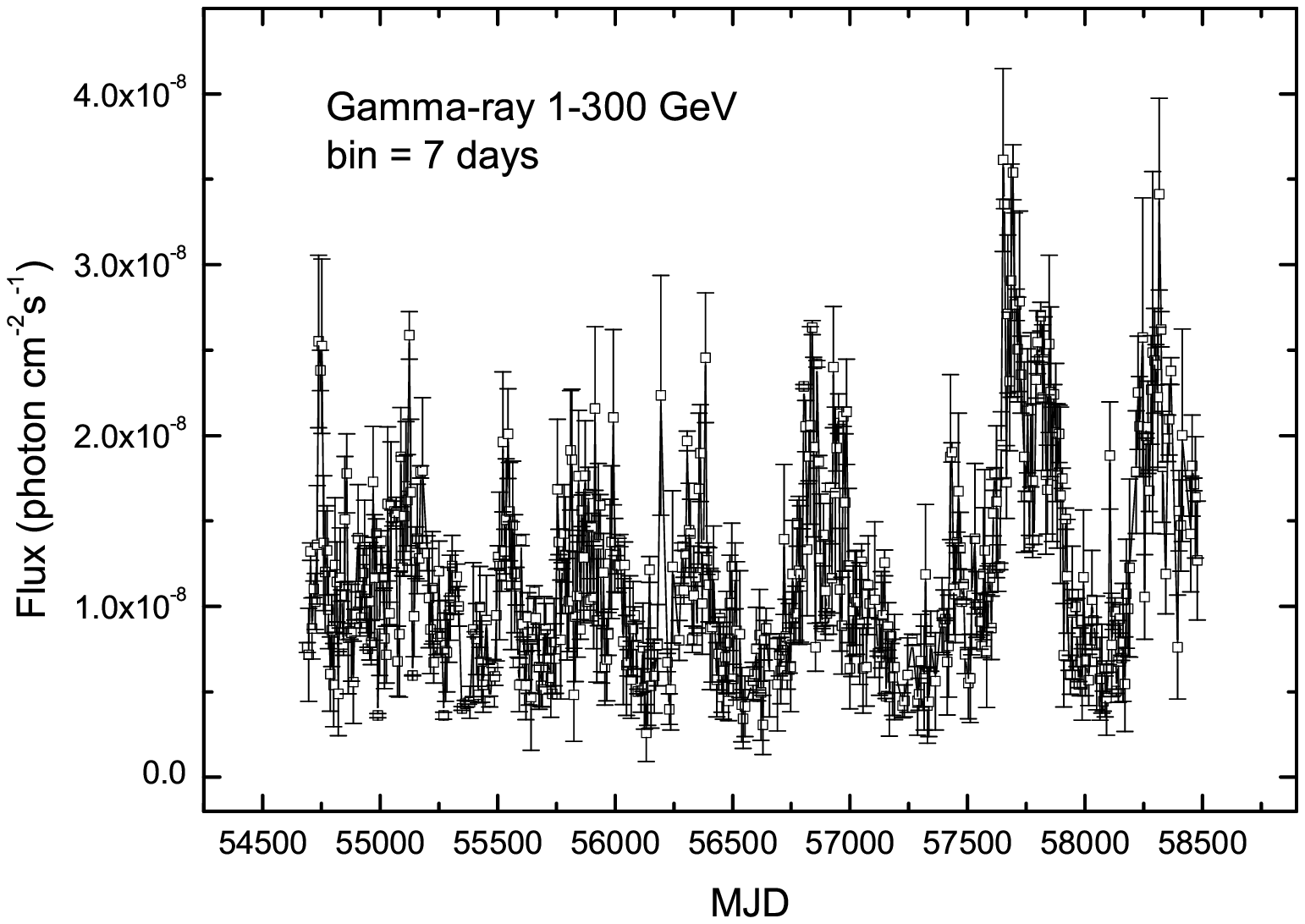}
\includegraphics[angle=0,scale=0.46]{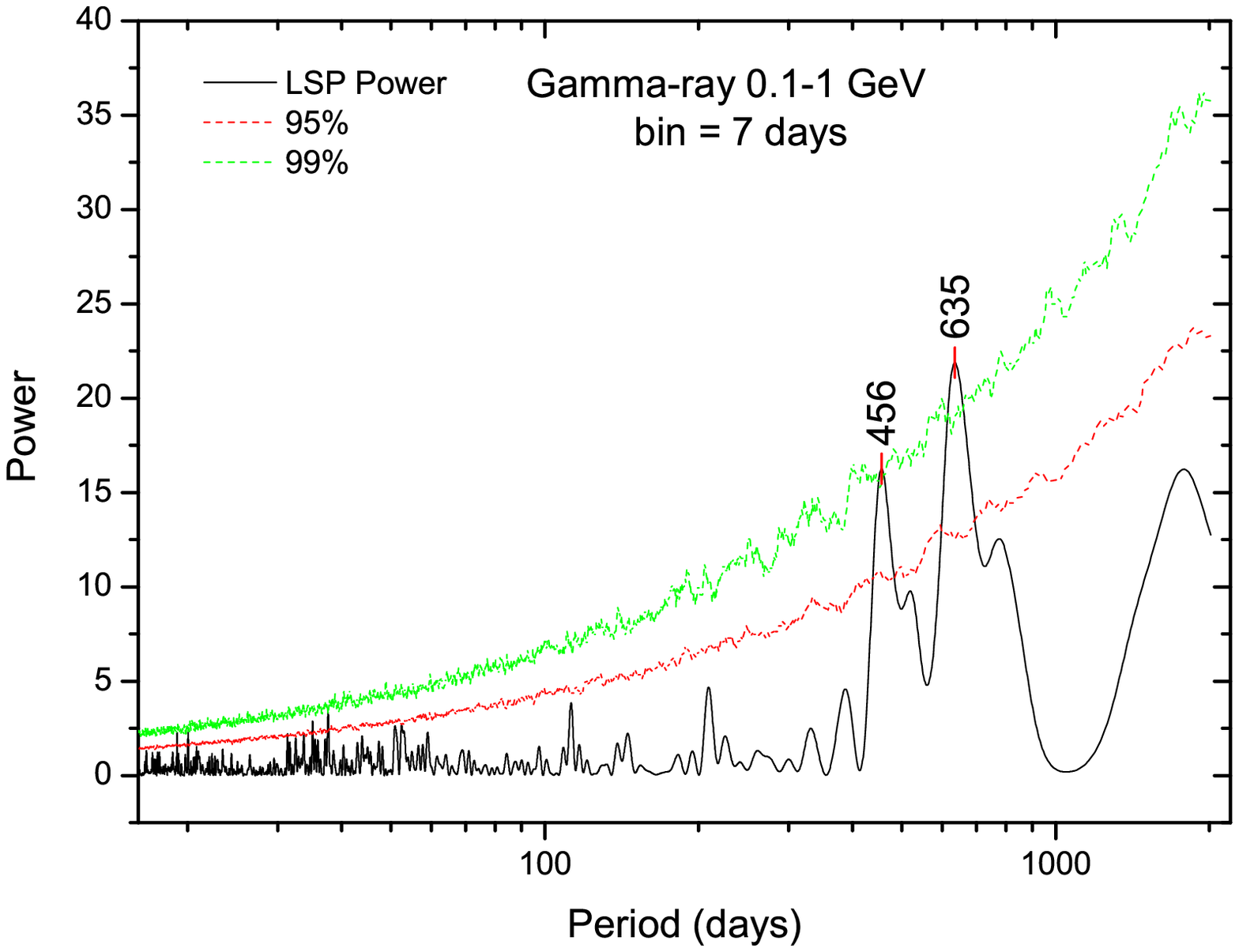}
\includegraphics[angle=0,scale=0.46]{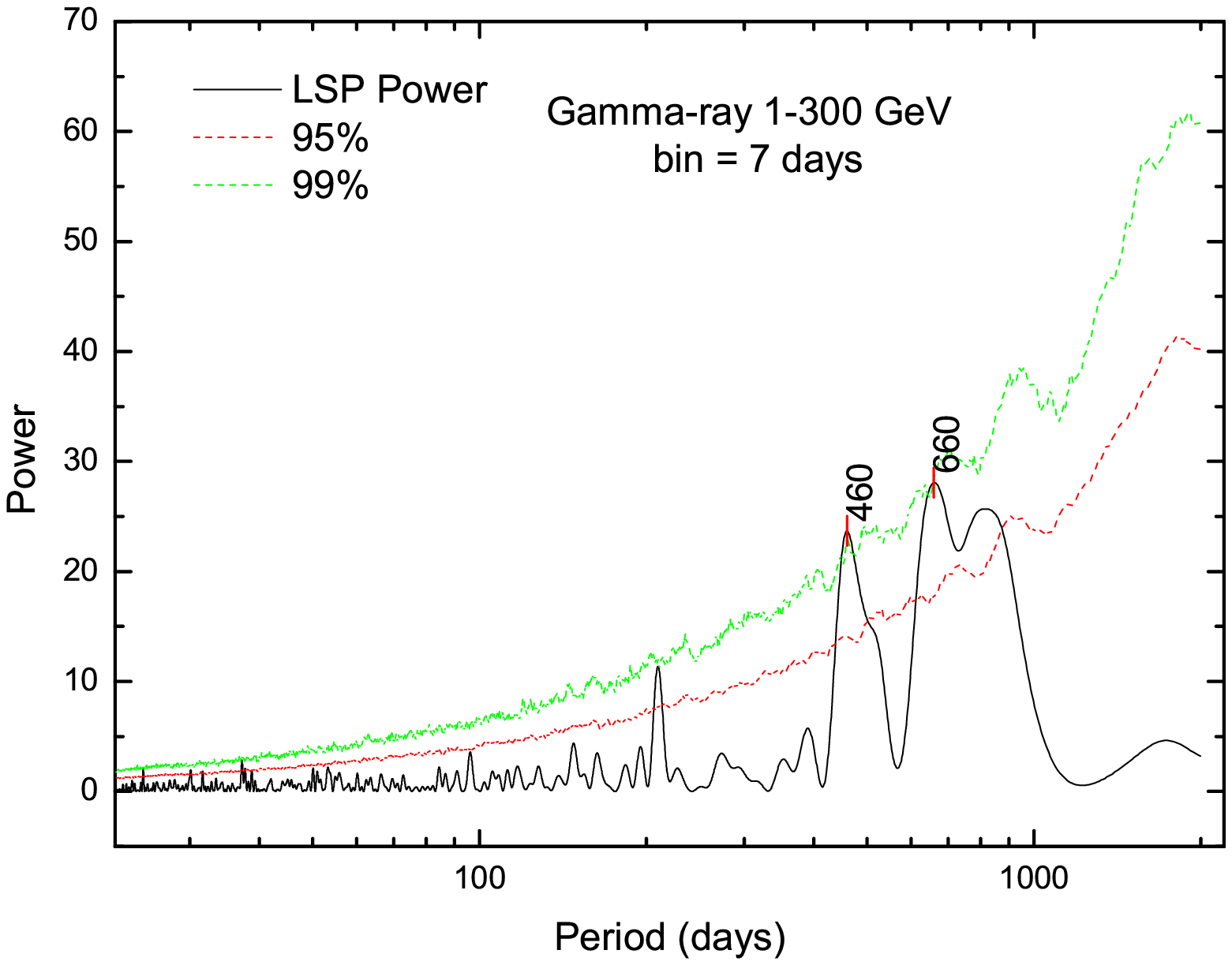}
\caption{The LSP results of $\rm\gamma$-Ray band (1.0-300 GeV) light curve with 7 days bin for 3FGL J0449.4-4350. The red and green dashed lines represent the confidence level of $95\%$, and $99\%$ respectively.  }
\end{figure}
%Periods in days corresponding to the relatively high peaks are marked. There was a 456 days periodicity with very high statistical significance.

\begin{figure}
\centering
\includegraphics[angle=0,scale=0.18]{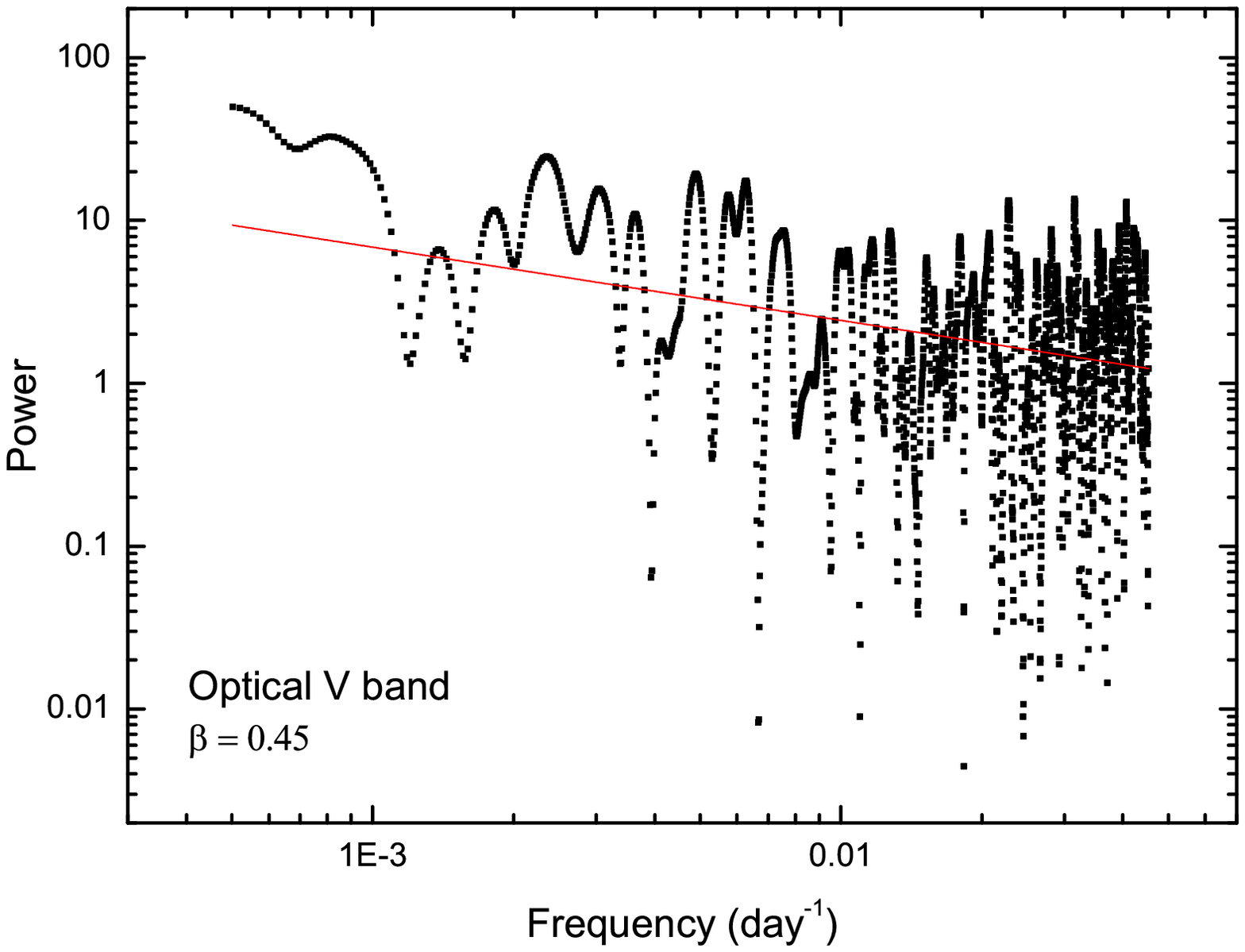}
\includegraphics[angle=0,scale=0.18]{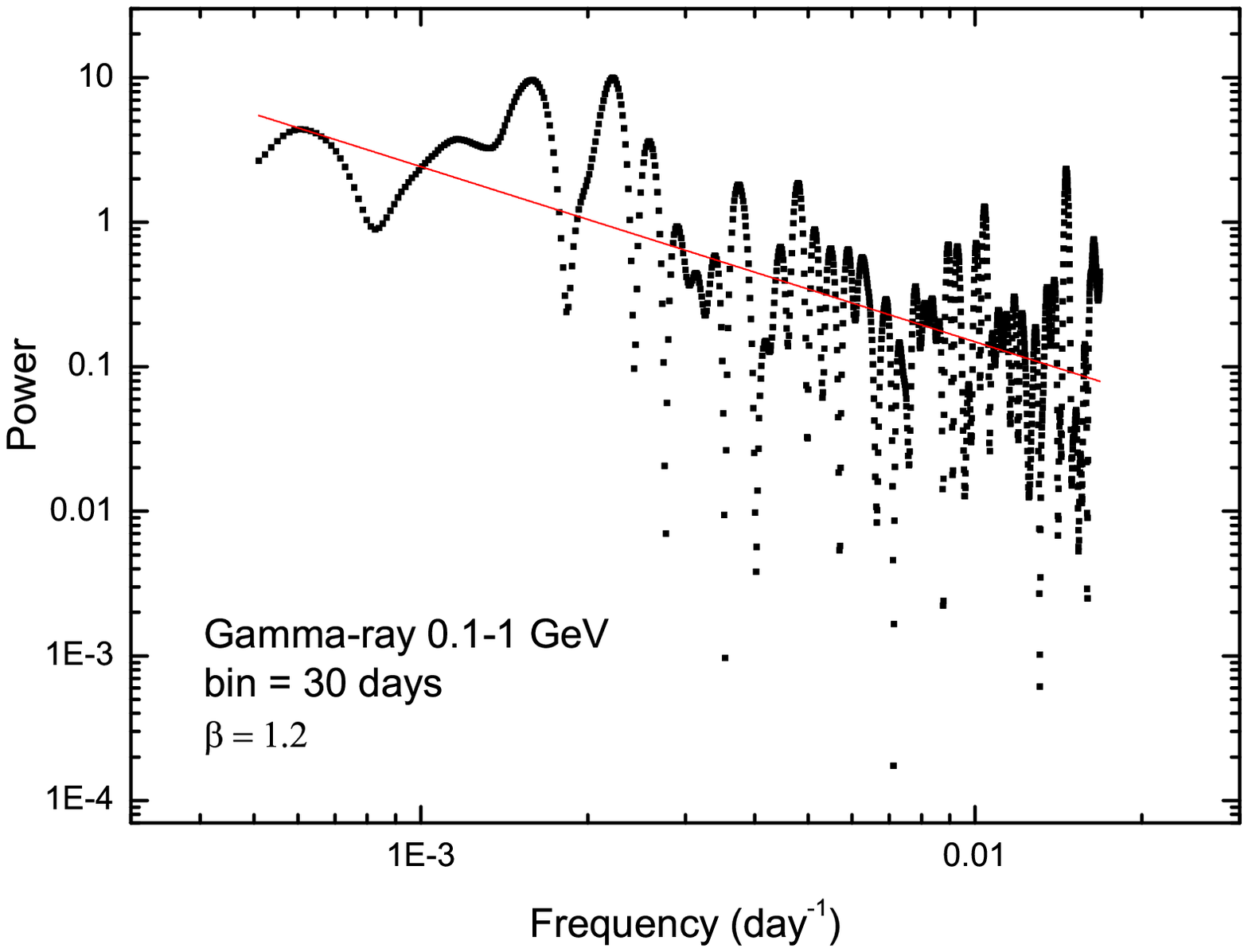}
\includegraphics[angle=0,scale=0.18]{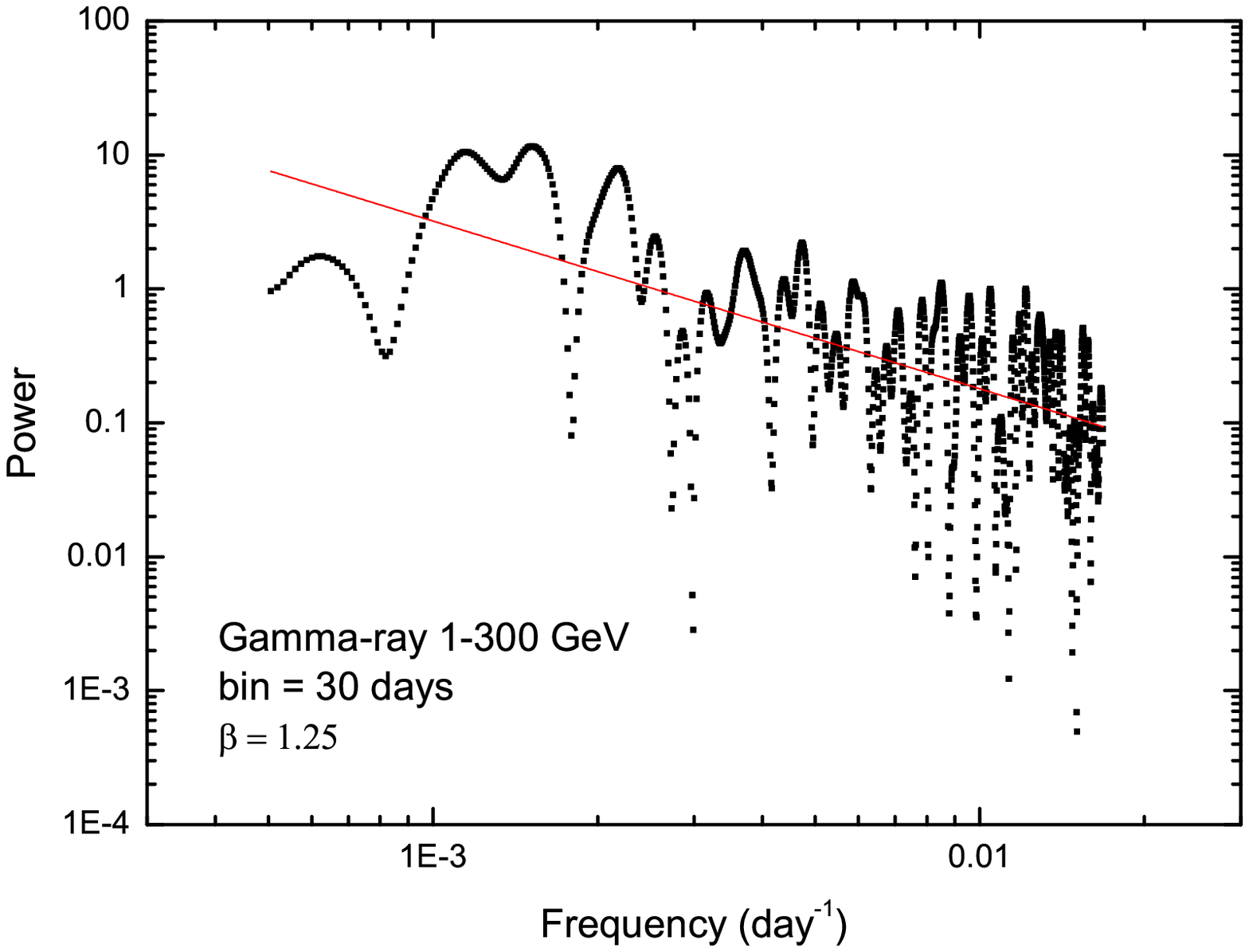}
\includegraphics[angle=0,scale=0.18]{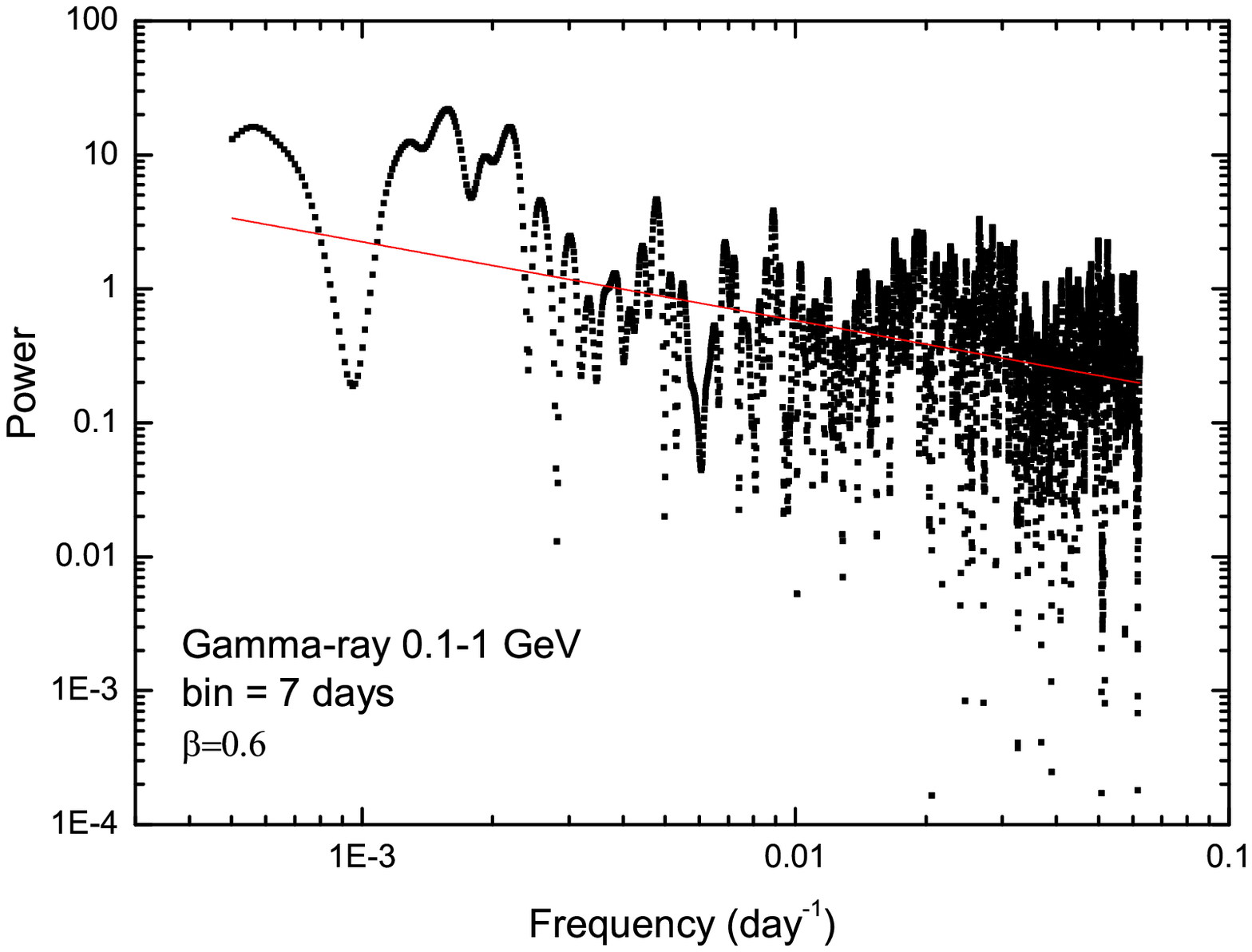}
\includegraphics[angle=0,scale=0.18]{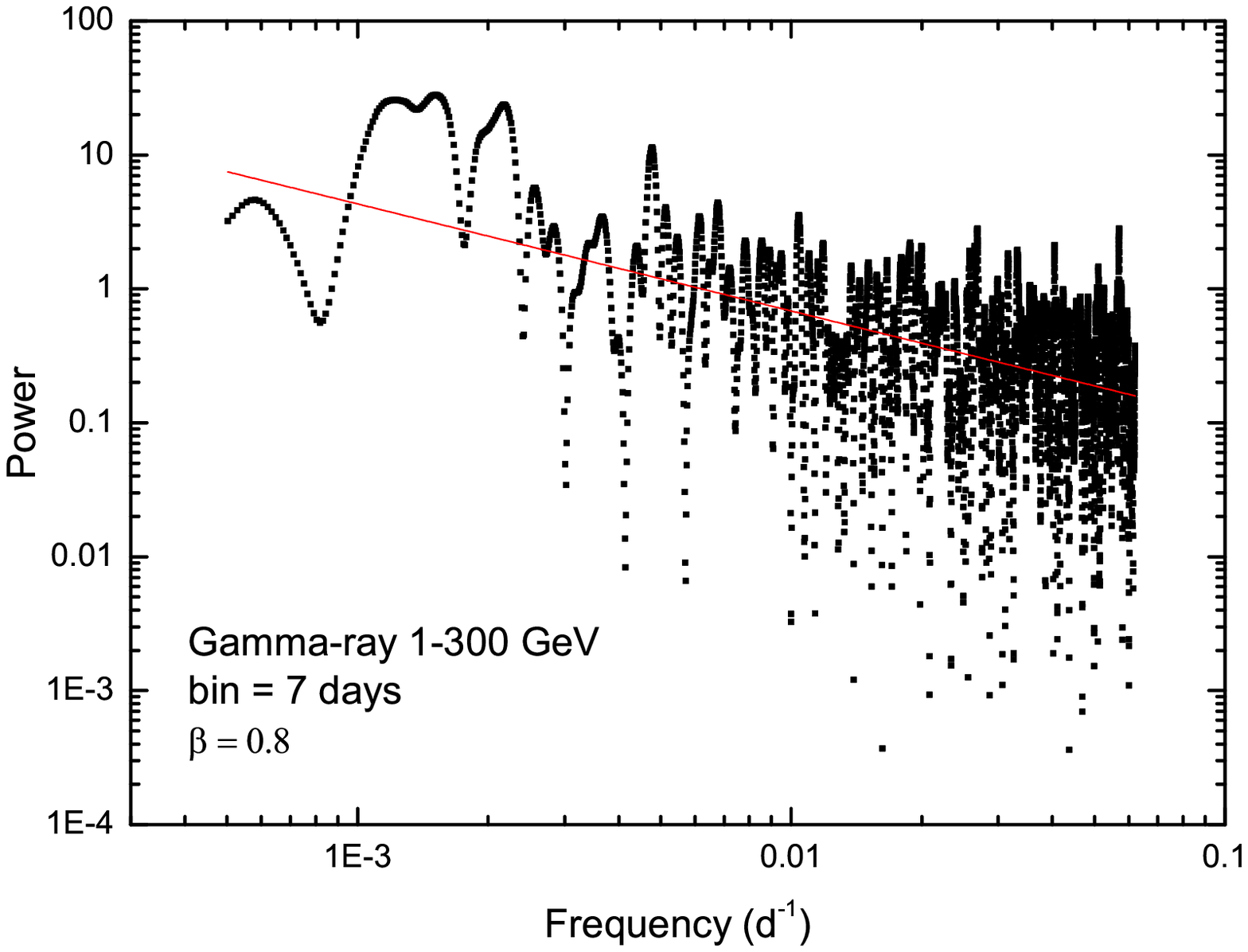}
\includegraphics[angle=0,scale=0.17]{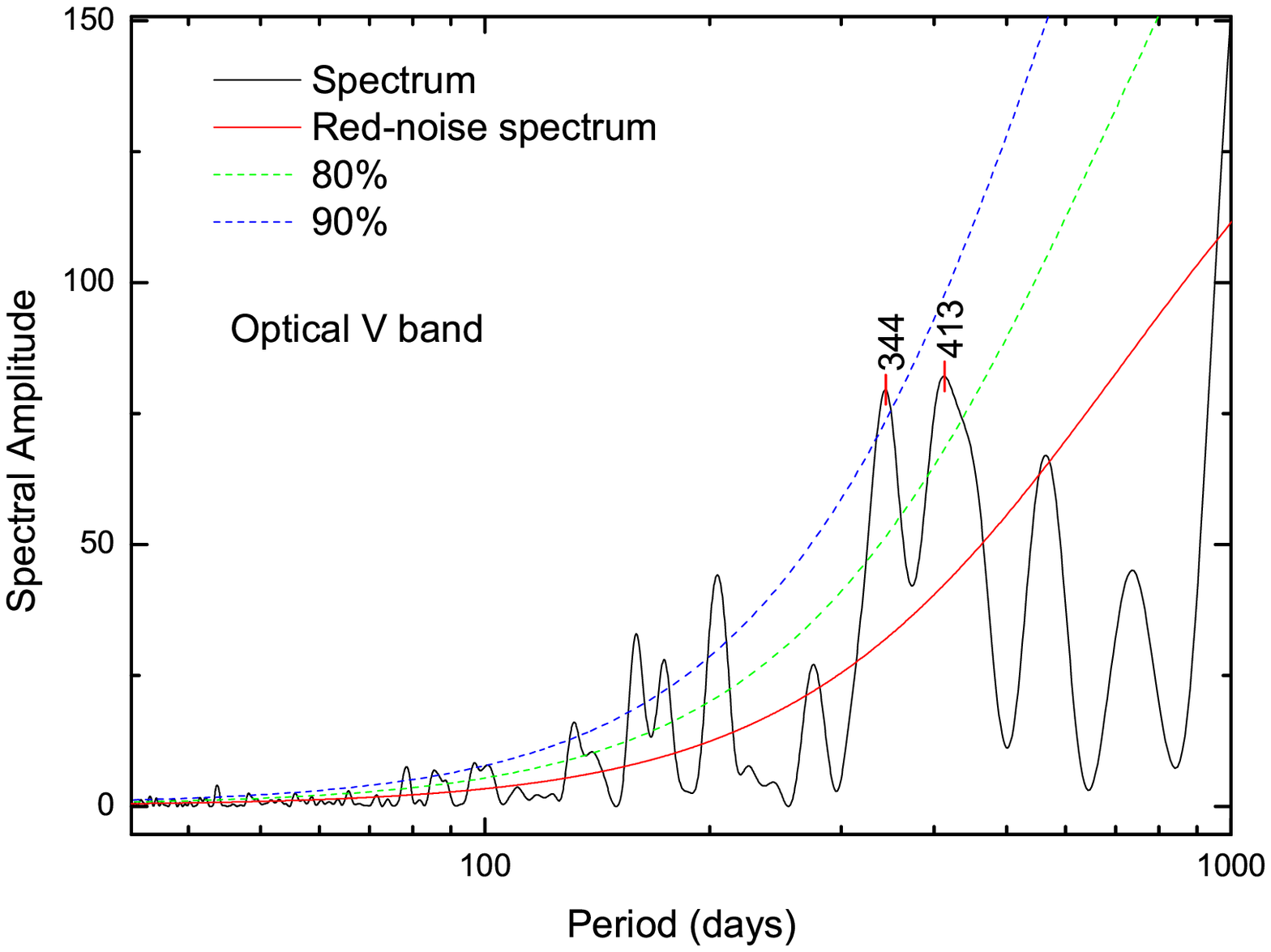}
\includegraphics[angle=0,scale=0.17]{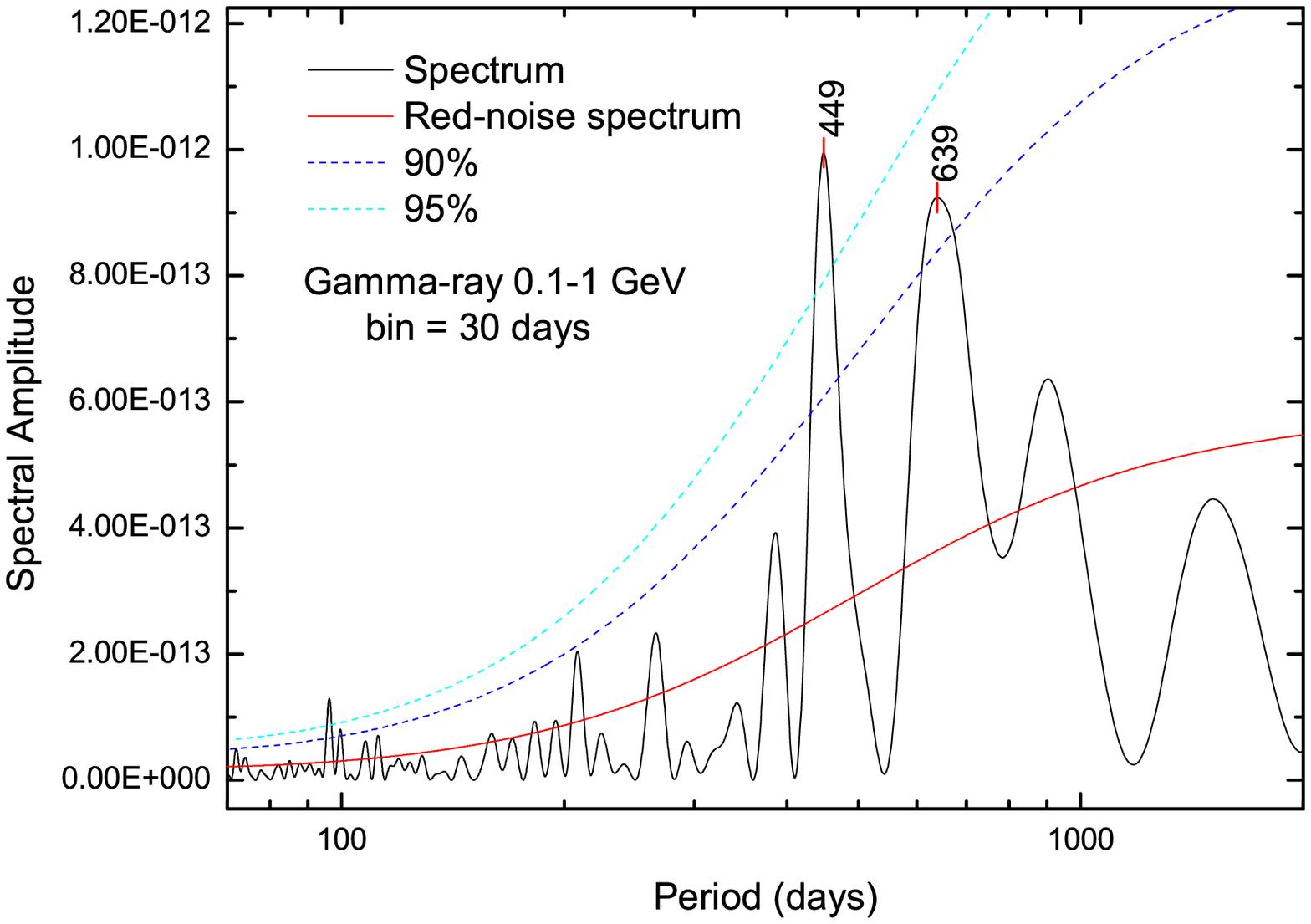}
\includegraphics[angle=0,scale=0.17]{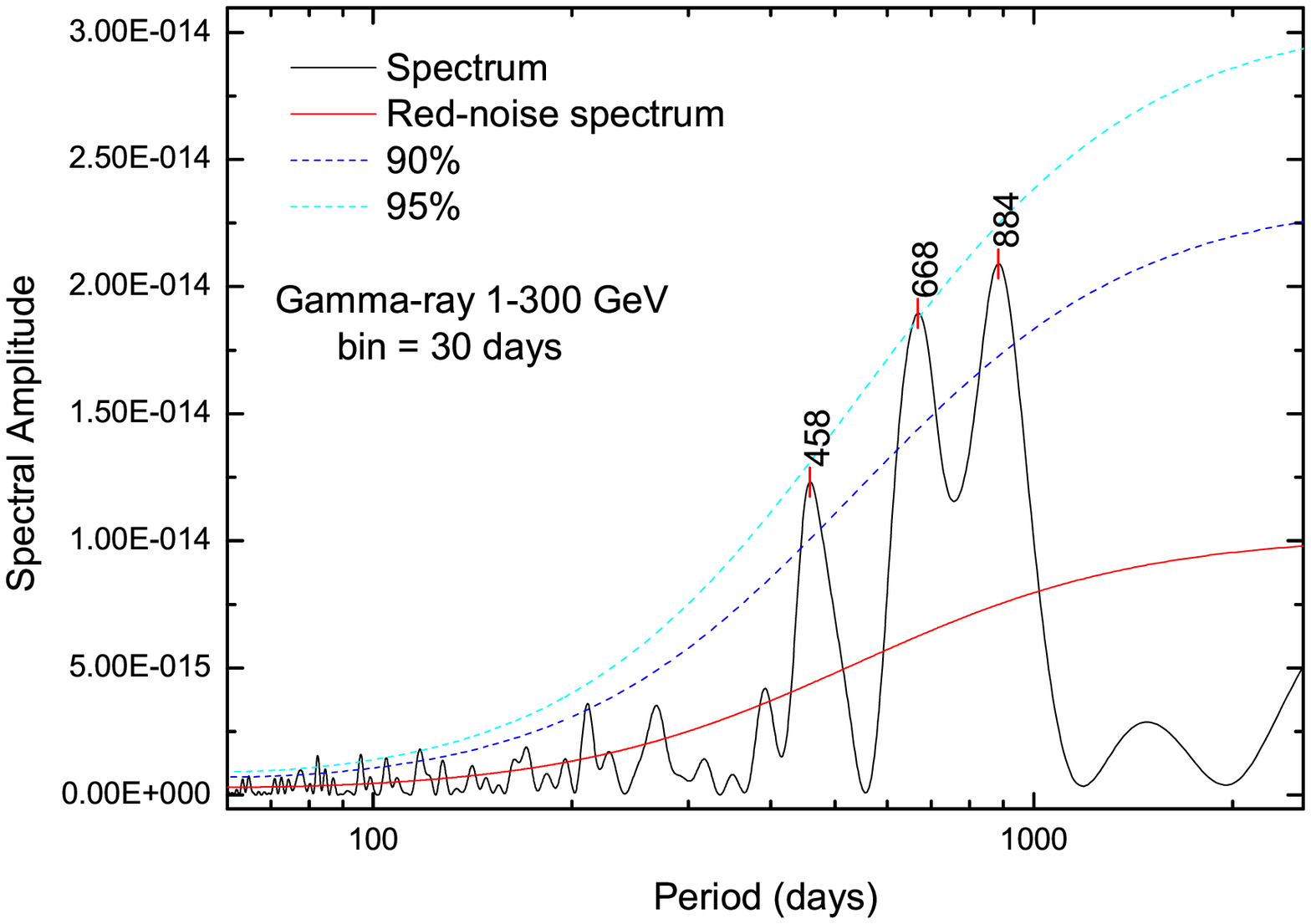}
\includegraphics[angle=0,scale=0.17]{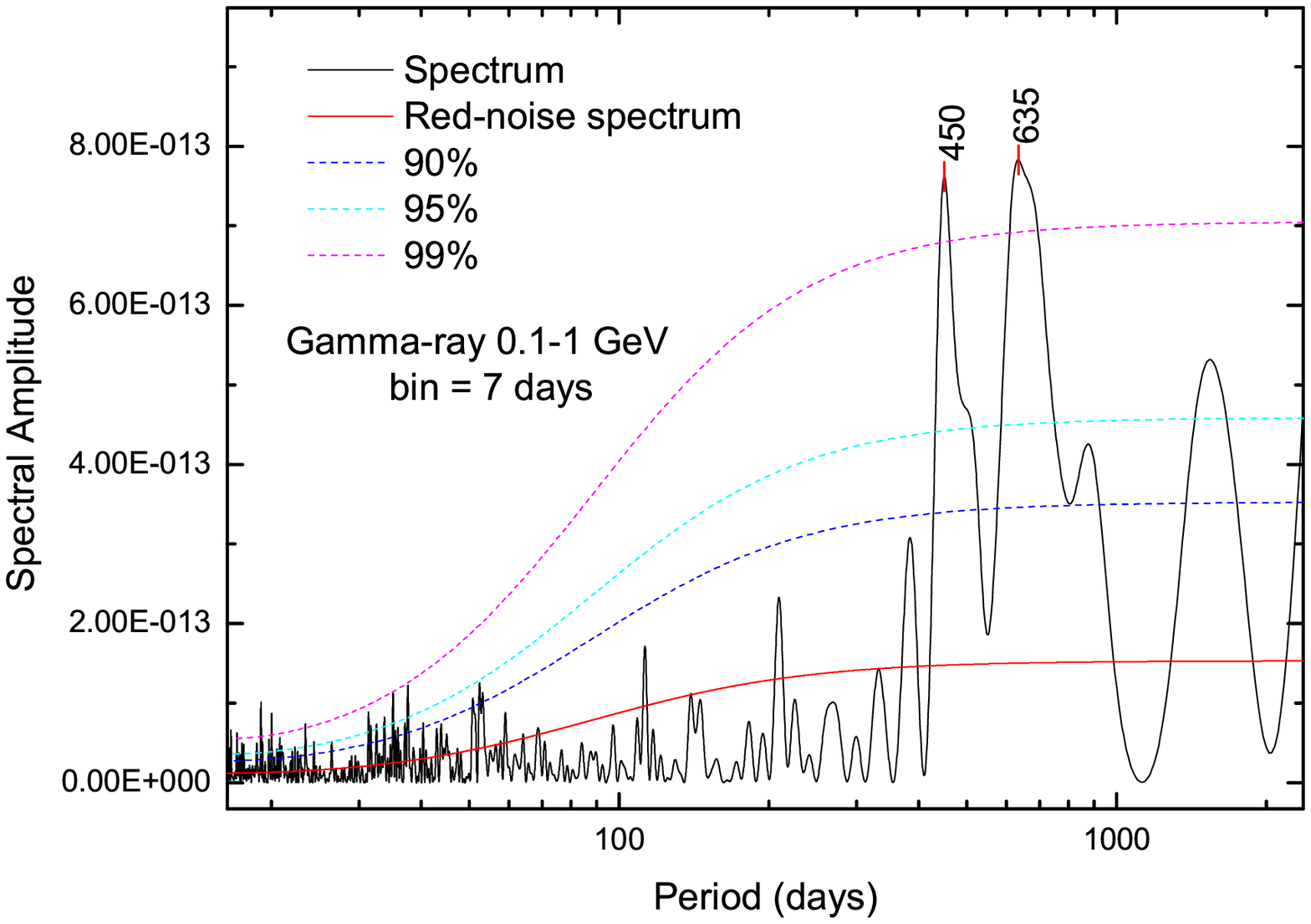}
\includegraphics[angle=0,scale=0.17]{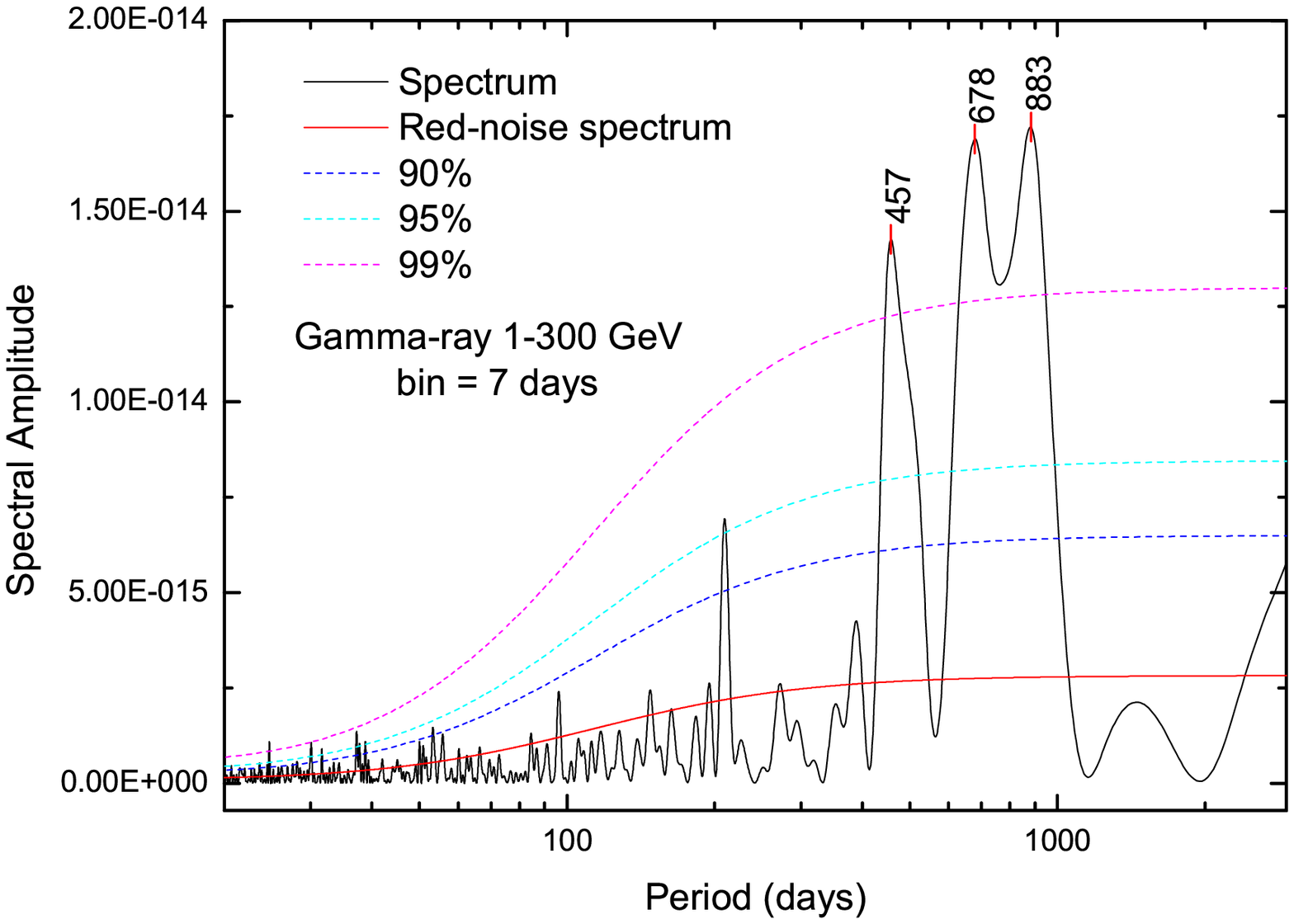}
\includegraphics[angle=0,scale=0.18]{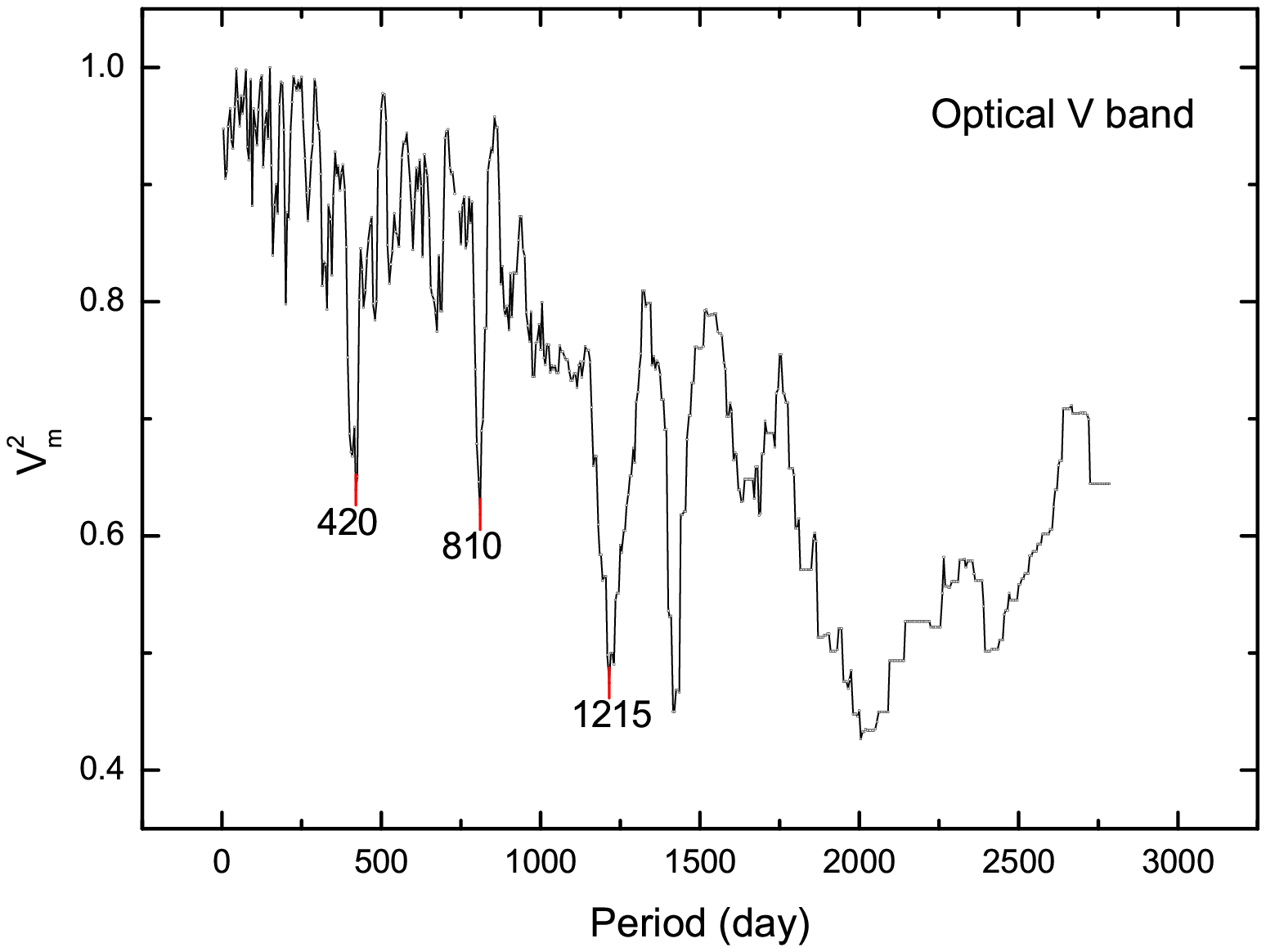}
\includegraphics[angle=0,scale=0.2]{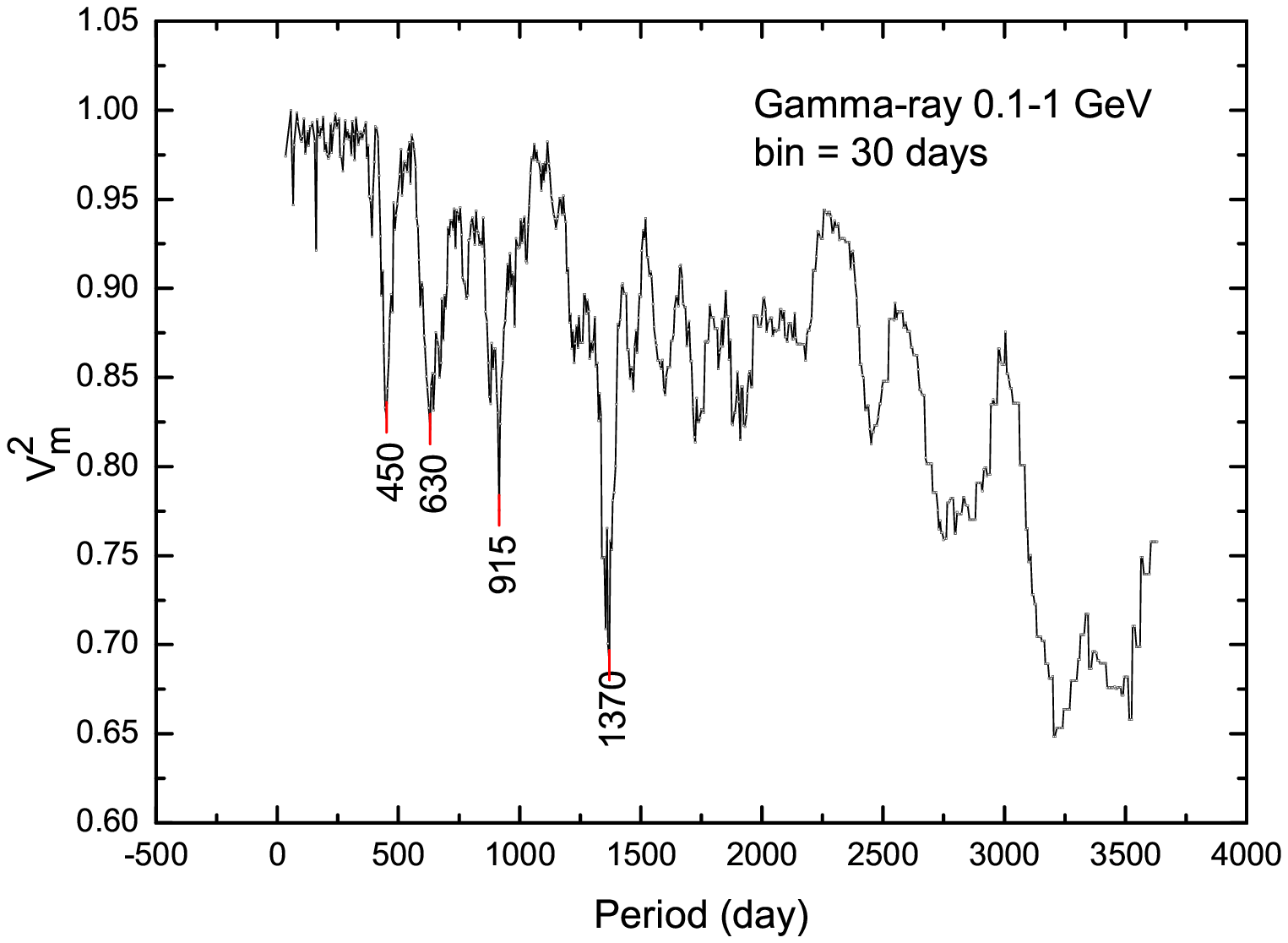}
\includegraphics[angle=0,scale=0.2]{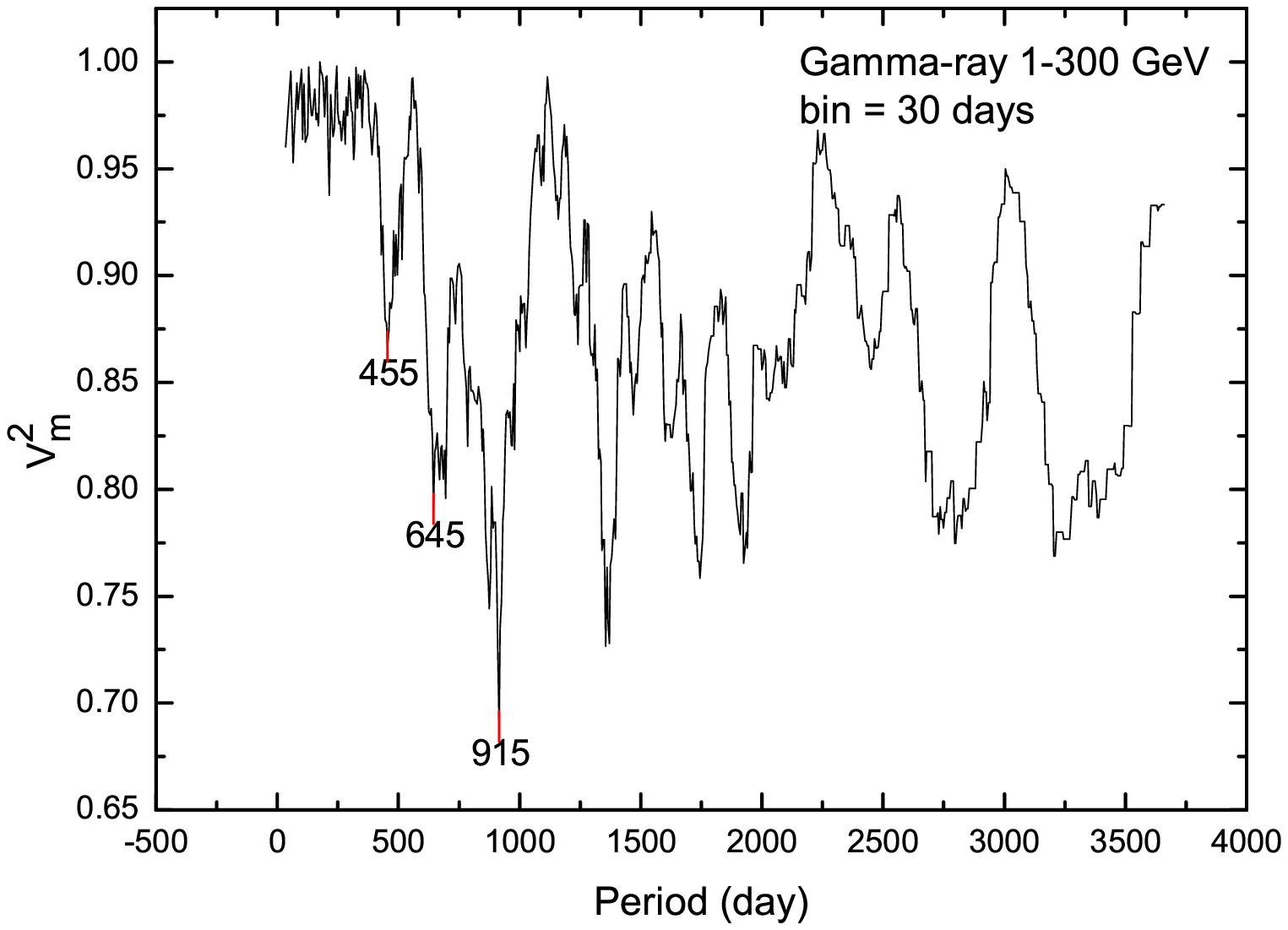}
\includegraphics[angle=0,scale=0.2]{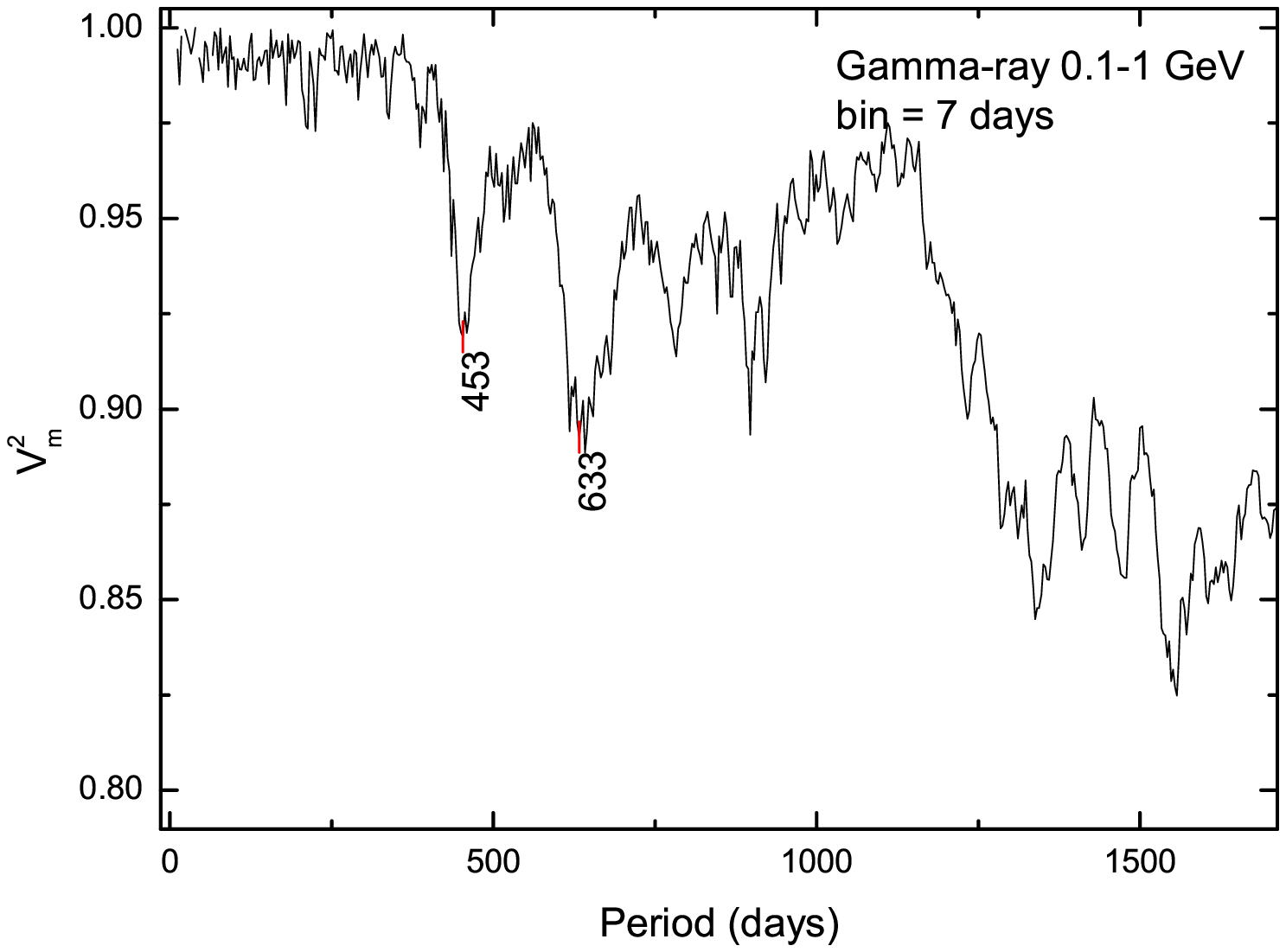}
\includegraphics[angle=0,scale=0.2]{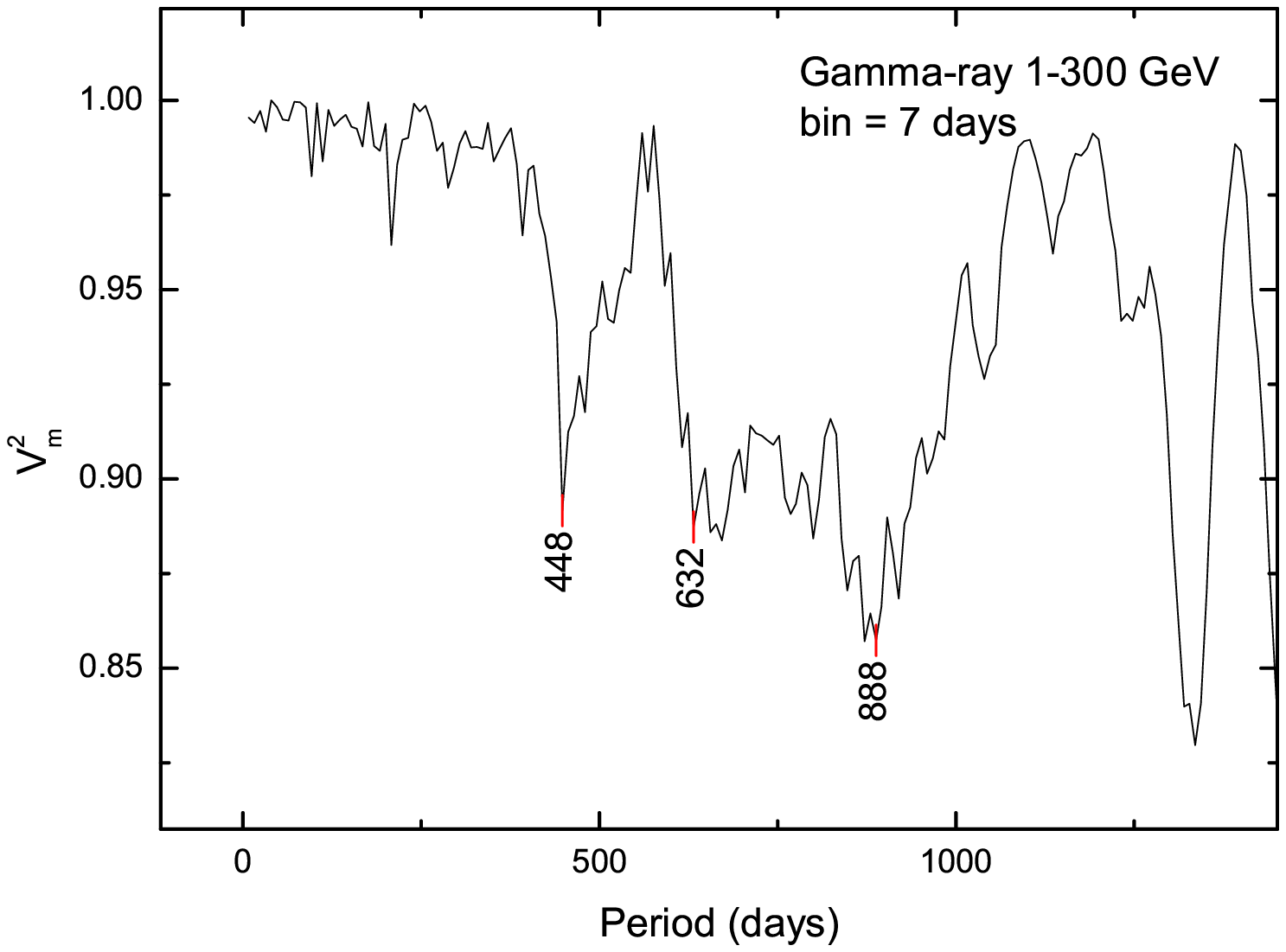}
\includegraphics[angle=0,scale=0.19]{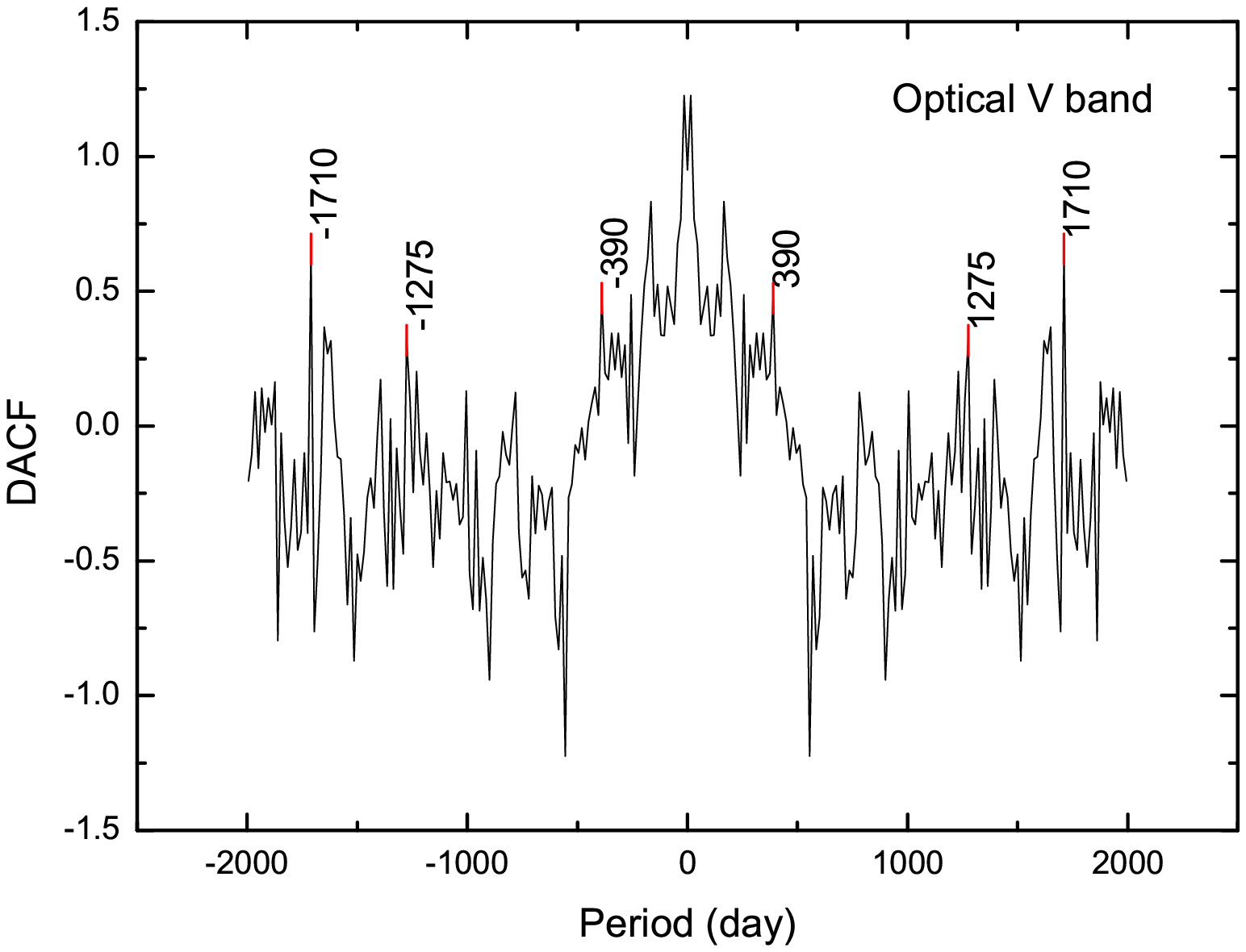}
\includegraphics[angle=0,scale=0.2]{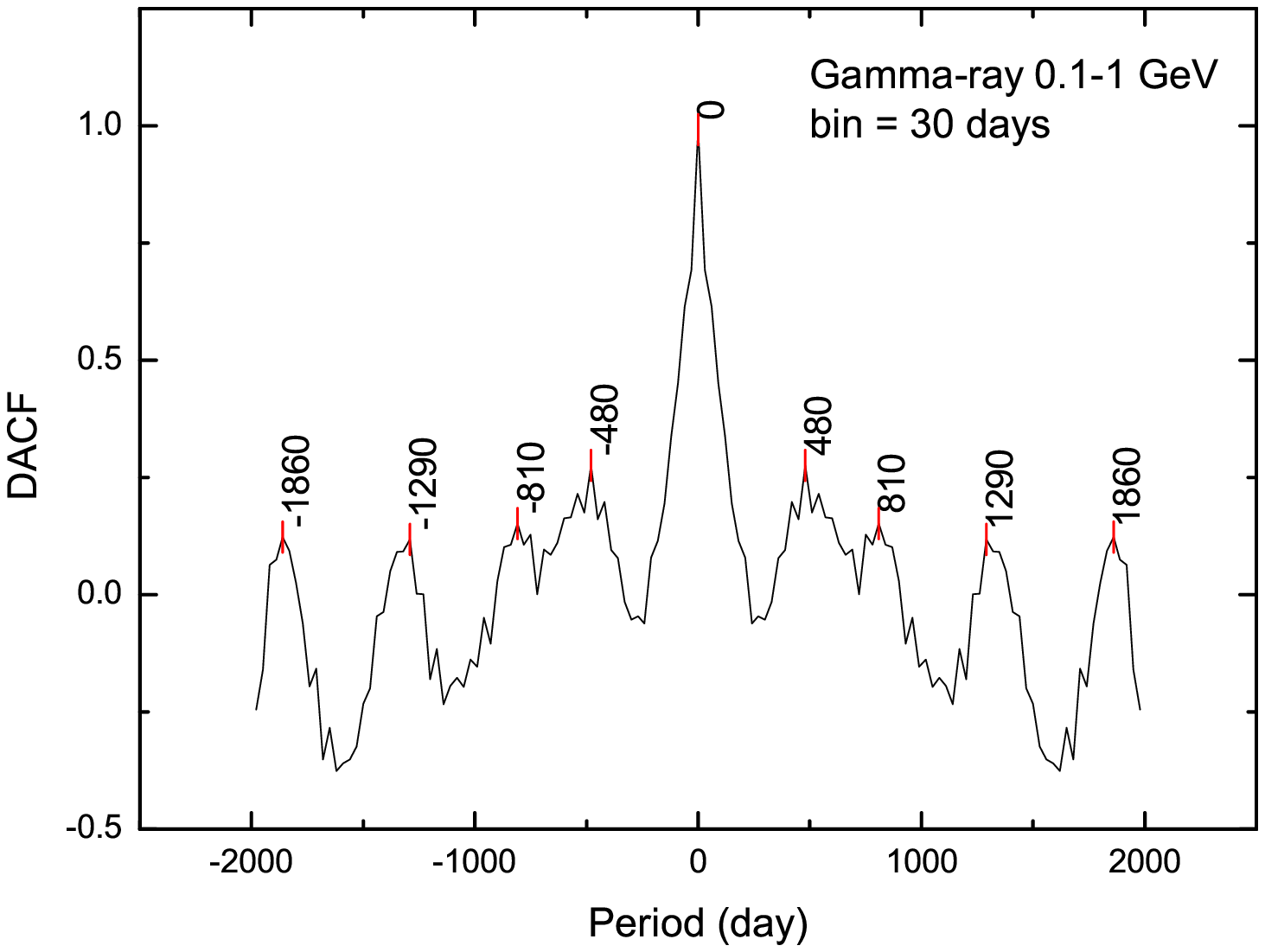}
\includegraphics[angle=0,scale=0.2]{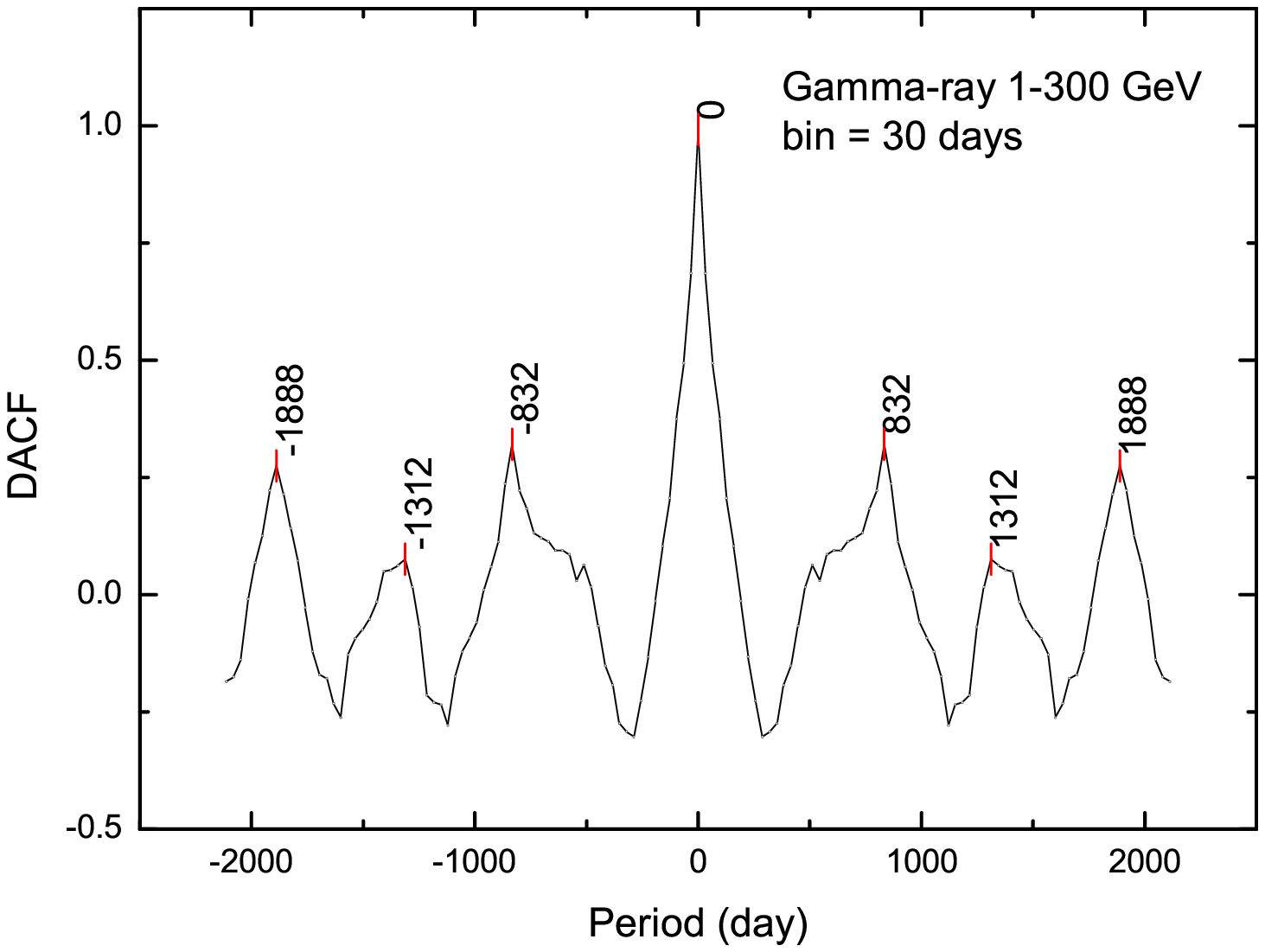}
\includegraphics[angle=0,scale=0.2]{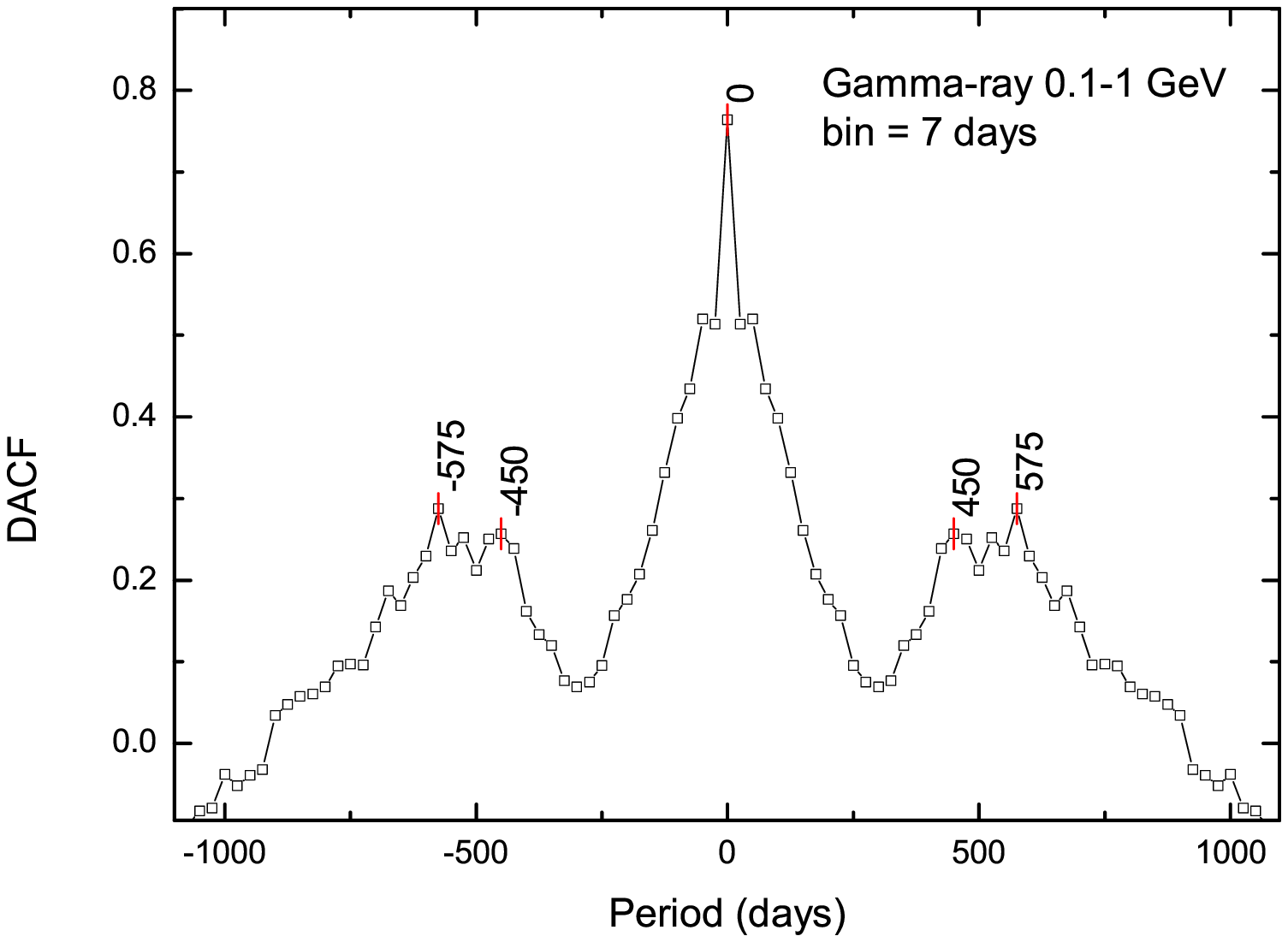}
\includegraphics[angle=0,scale=0.18]{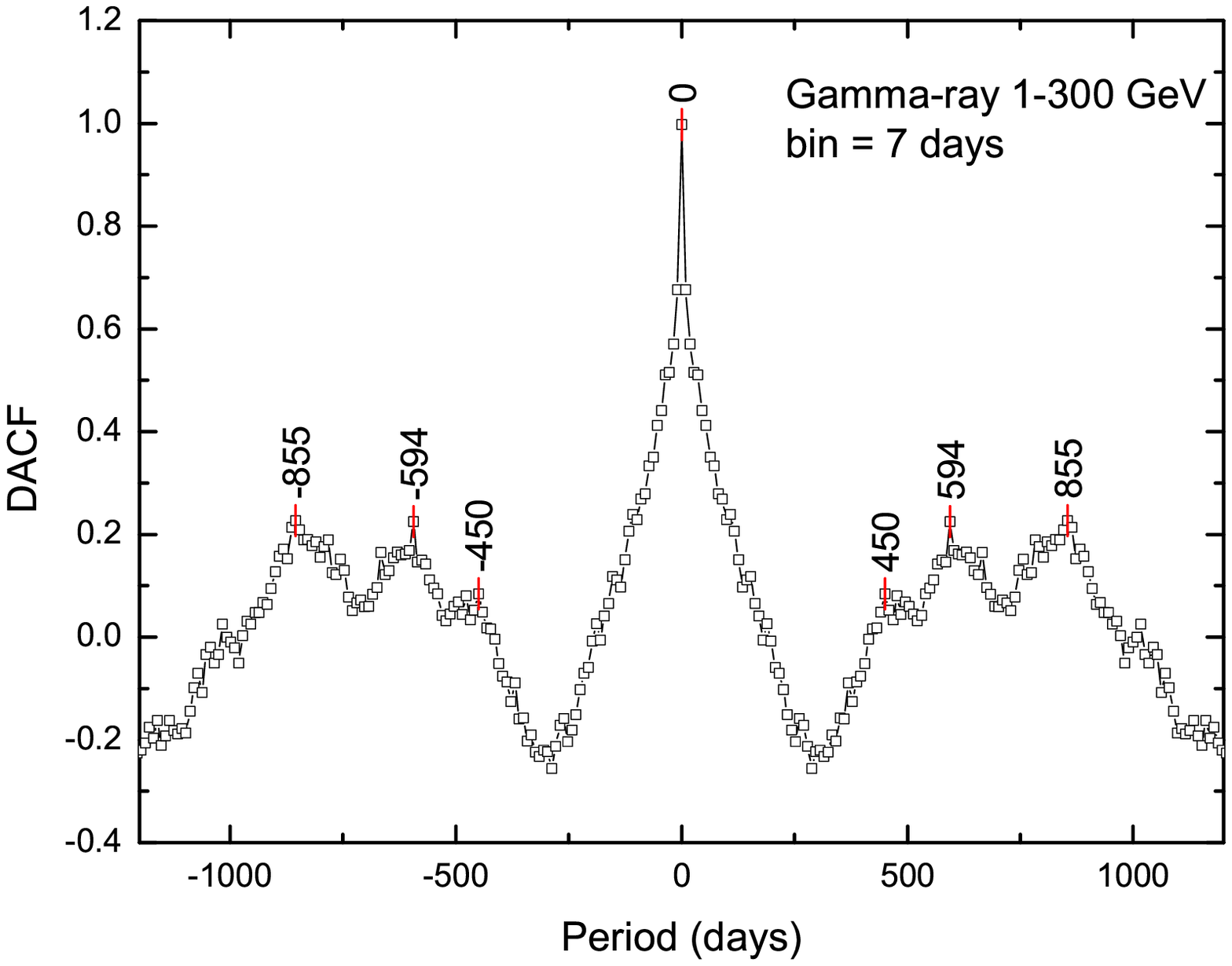}
\caption{First line panels, the broad power spectrum, fitting with $p(f)=f^{-\beta}$, for optical and $\gamma$-ray band with 30 and 7 days bins for 3FGL J0449.4-4350. second line panels, the Redfit38 results of the 5 wavelength bands. the third and forth line panels, the J-K and DACF method results of the 5 wavelength bands, respectively.}
%, 480:450:$\approx$1:1, 633:1275:1290:1312:757:594$\approx$1:2:2:2:1:1. These proportional relationships indicate that $T1=450$ and $T2=630$ days are two more significant quasi-period of 3FGL 0449.4+4350.}
\end{figure}

\begin{deluxetable}{rrrrrrrrrr}
\tabletypesize{\scriptsize}
\centering
\tablecaption{Summary of the Results of the Periodicity Analysis in Multi-wavelength bands for 3FGL J0449.4-4350}
\tablewidth{435pt}
\tablehead{Bands &LSP&Conf.&FAP& J-K & \colhead{$f$}&   REDFIT38  & \colhead{Conf.} & DACF  & Coef. \\
\ &(days)&&&(days)&     & \colhead{(days)}& & \colhead{(days)} &
}
\startdata
            &&&&&&&&\\
Optical V-band &425$\pm$40&$\approx99.7\%$&3.4E-09&420$\pm$17     & 0.56      &344$\pm$25  &$\approx90\%$   &390$\pm$90    &0.48\\
           &&&&810$\pm$19     & 0.62      &413$\pm$33  & $\approx85\%$  &1275$\pm$22    &0.32\\
           &&&&1215$\pm$37    & 1.11      &            &           &1710$\pm$12  &0.66\\

Fermi LAT   &450$\pm$23&$\approx99.7\%$ &6.3E-03&450$\pm$25     & 0.21      &449$\pm$24  &$\approx98\%$         &480$\pm$94    &0.28\\
0.1-1.0 GeV &630$\pm$58&$\approx98\%$ &8.8E-03&630$\pm$20     & 0.22      &639$\pm$77  &$\approx90\%$     &810$\pm$129     &0.15\\
30 days bin &&&&915$\pm$29     & 0.29      &            &                  &1290$\pm$124    &0.12\\
            &&&&1370$\pm$21    & 0.45  &&&&\\

Fermi LAT  &459$\pm$33&$\approx99\%$&4.7E-02&455$\pm$29     & 0.15     &458$\pm$35  &$\approx94\%$     &832$\pm$105    &0.32\\
1.0-300 GeV&658$\pm$70&$\approx99\%$&1.3E-03&645$\pm$22     & 0.26     &668$\pm$74  &$\approx95\%$    &1312$\pm$159   &0.08\\
30 days bin&&&&915$\pm$14     & 0.45     &884$\pm$120 & $\approx94\%$  & &\\

           & &&&&&&&&\\
Fermi LAT   &456$\pm$31&$\approx99\%$&4.9E-05&453$\pm$23     & 0.09      &450$\pm$93  &$\gg99\%$         &450$\pm$34    &0.26\\
0.1-1.0 GeV &635$\pm$65&$\approx99\%$&1.8E-07&633$\pm$61     & 0.12      &635$\pm$192  &$\gg99\%$     &575$\pm$ 30    &0.29\\
7 days bin   && &  &   &      &  &                &     & \\

           & &&&&&&&&\\
Fermi LAT  &460$\pm$46&$\approx99\%$&2.9E-08&448$\pm$25     & 0.14     &457$\pm$46  &$\gg99\%$     &450$\pm$55    &0.08\\
1.0-300 GeV&660$\pm$61&$\approx99\%$&3.6E-10&632$\pm$21     & 0.14     &678$\pm$75  &$\gg99\%$   &594$\pm$60  &0.22\\
7 days bin&&&&888$\pm$54      & 0.18     &883$\pm$111   & $\gg99\%$  &855$\pm$59 &0.23 \\
        &  &&&&&&&&\\
%\hline
\enddata
\tablecomments{The data of each band for the four periodic analysis methods(LSP, J-K, redfit38, DACF methods, as shown in Fig. 2, Fig. 3, and Fig. 4). The uncertainties are half with at half maximum (HWHM) of the peaks}
\end{deluxetable}

The REDFIT38 program\footnote[2]{https://www.geo.uni-bremen.de/geomod/staff/mschulz/\#software2} (Schulz and Mudelsee 2002) based on the LSP (Lomb 1976; Scargle 1982) often be performed to estimated the red noise level in the light curve of blazars (Sandrinelli et al. 2016A, 2017, 2018; Gupta et al. 2019). In this program model the red noise as due to a first-order autoregressive (AR1) process.
The results of the REDFIT38 program by using the input parameter ($n_{50}=1$, $i_{win} = 0$, i.e. rectangular, as described by Schulz and Mudelsee 2002), are shown in Fig. 4. The dash lines in the second line panel of Fig. 4 denote the confidence levels.
In Fig. 4, a possible QPO of $449\pm35$ days ($\approx T1$) may be in 0.1-1 GeV $\gamma$-ray band for 30 days bin, with $99.7\%$ confidence level and similar periods in the other $\gamma$-ray band with low confidence levels, but for optical V-band  $413$ days peak with only about $85\%$. This maybe due to points of the light curve in optical V-band are too clustered to fit with AR1. As the author warned, the REDFIT38 program should not be used as black-box tool without checking the structure of a time series prior to its analysis (Schulz \& Mudelsee 2002). So the QPO is just a hint of a detection.
% The prominent peaks indicate the most possible periods in the corresponding light curve.

\begin{figure}
\centering
\includegraphics[angle=0,scale=0.5]{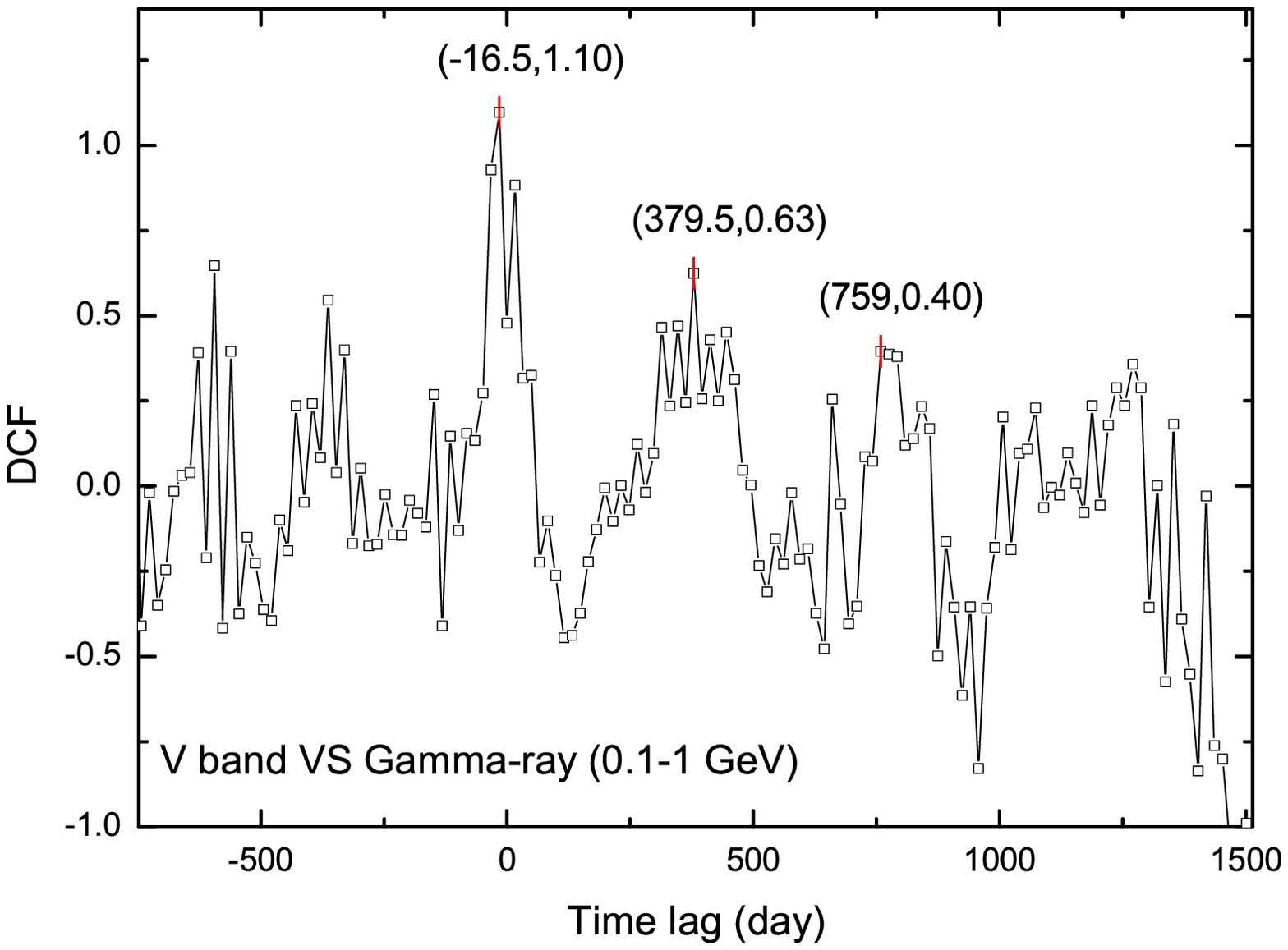}
\includegraphics[angle=0,scale=0.5]{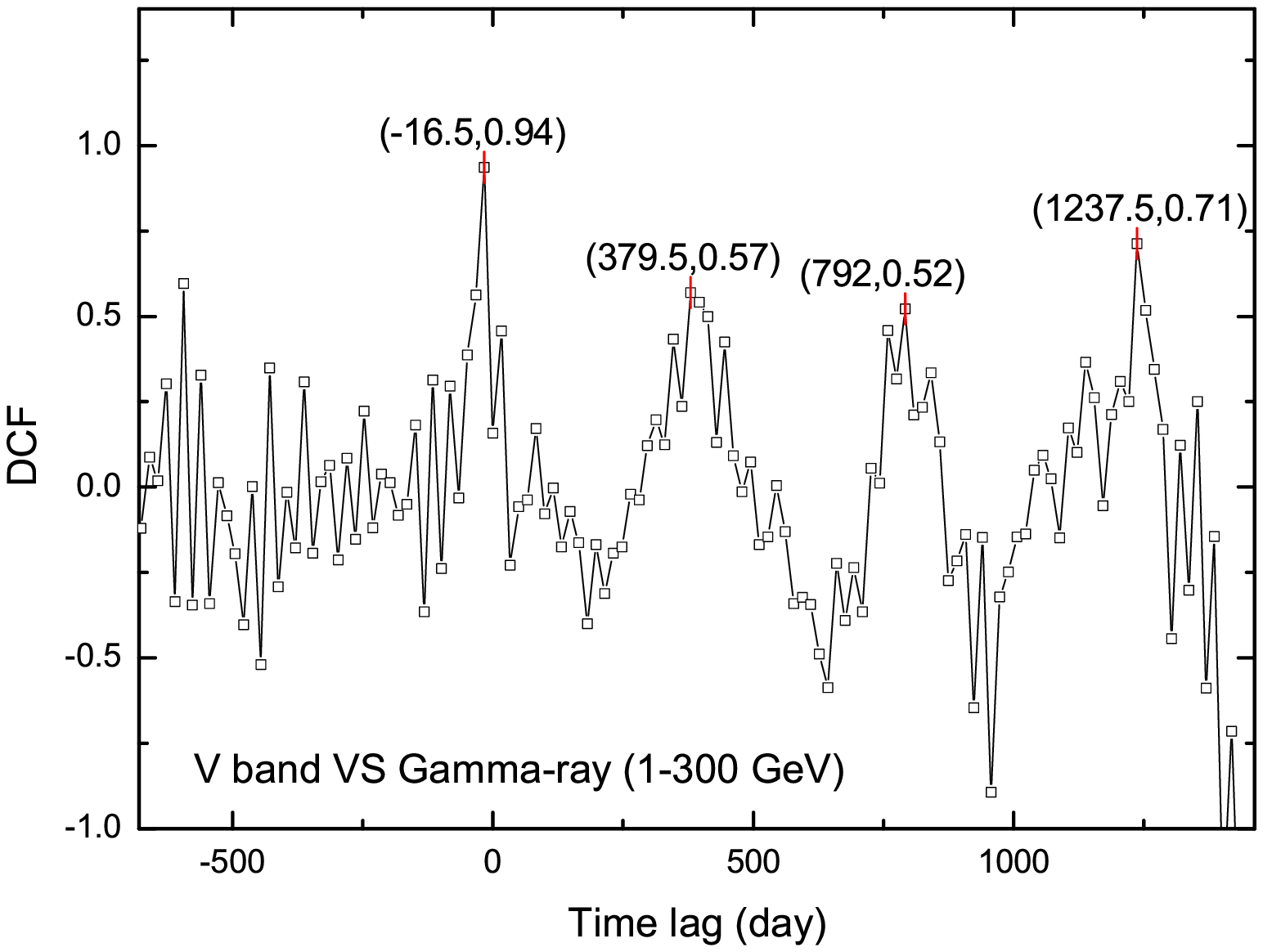}
\caption{The Discrete Correlation Function (DCF) results of $\gamma$-ray with 30 days bin and optical band for 3FGL J0449.4-4350.}
\end{figure}

\begin{figure}
\centering
\includegraphics[angle=0,scale=0.5]{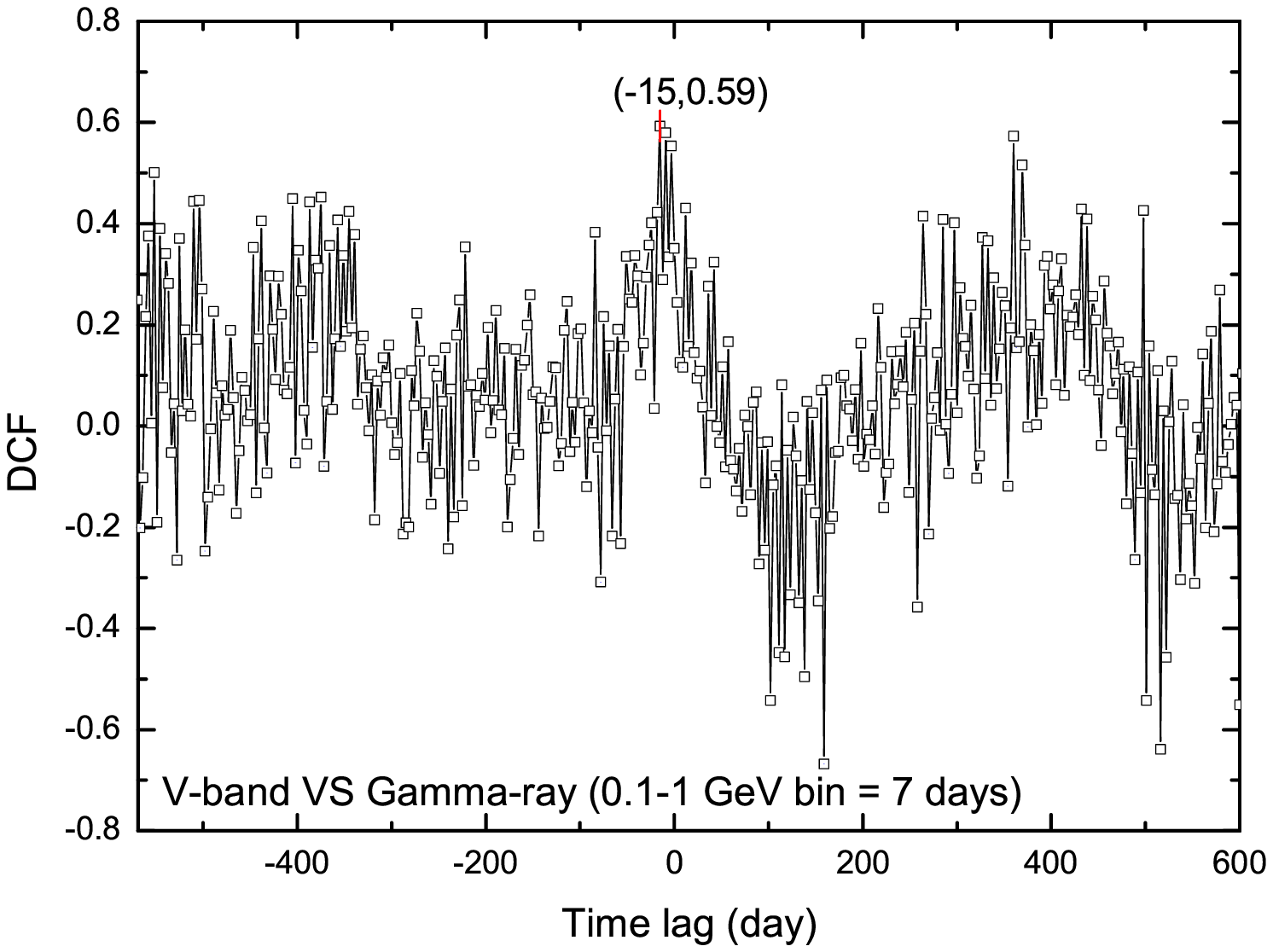}
\includegraphics[angle=0,scale=0.5]{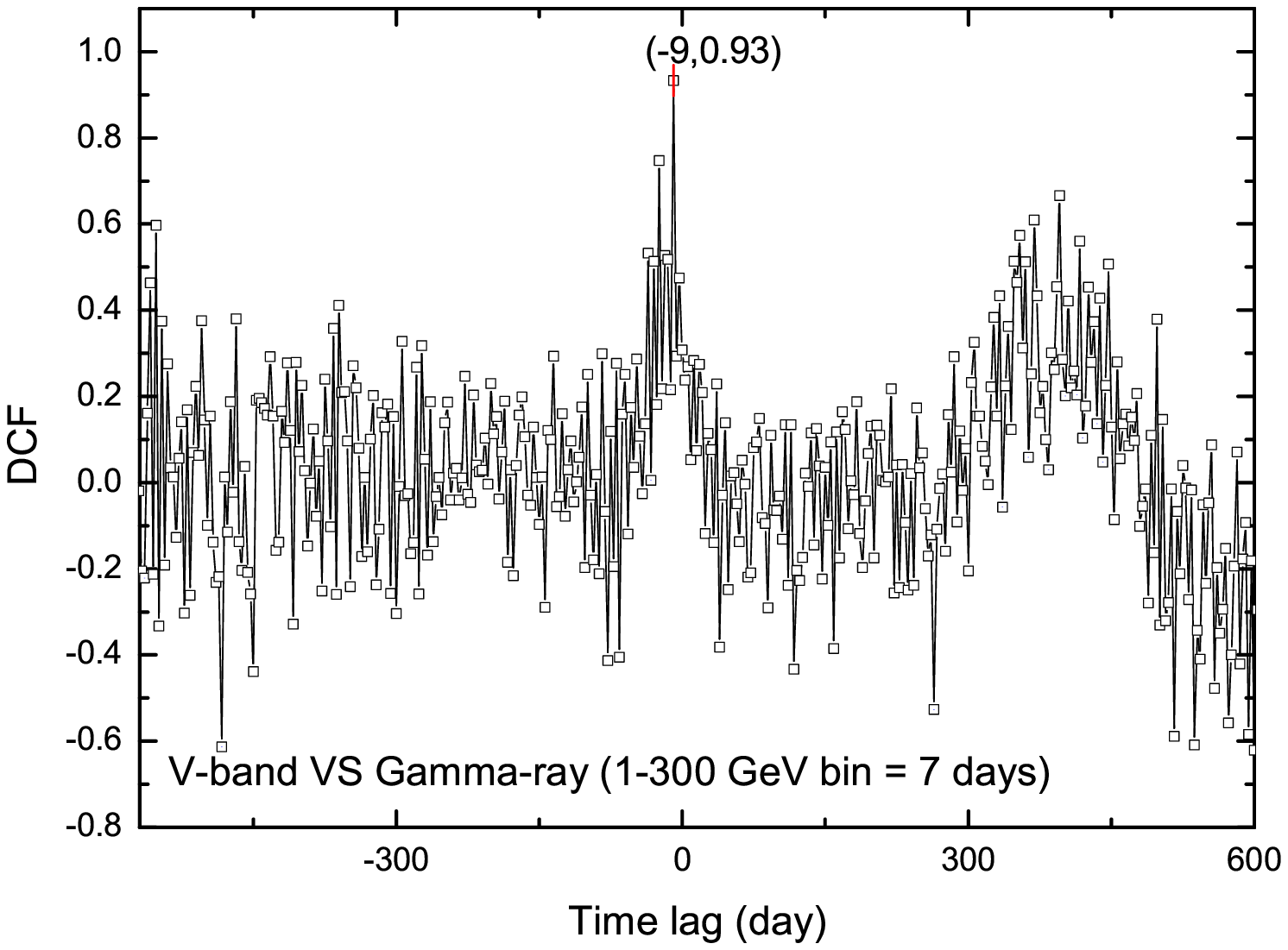}
\includegraphics[angle=0,scale=0.5]{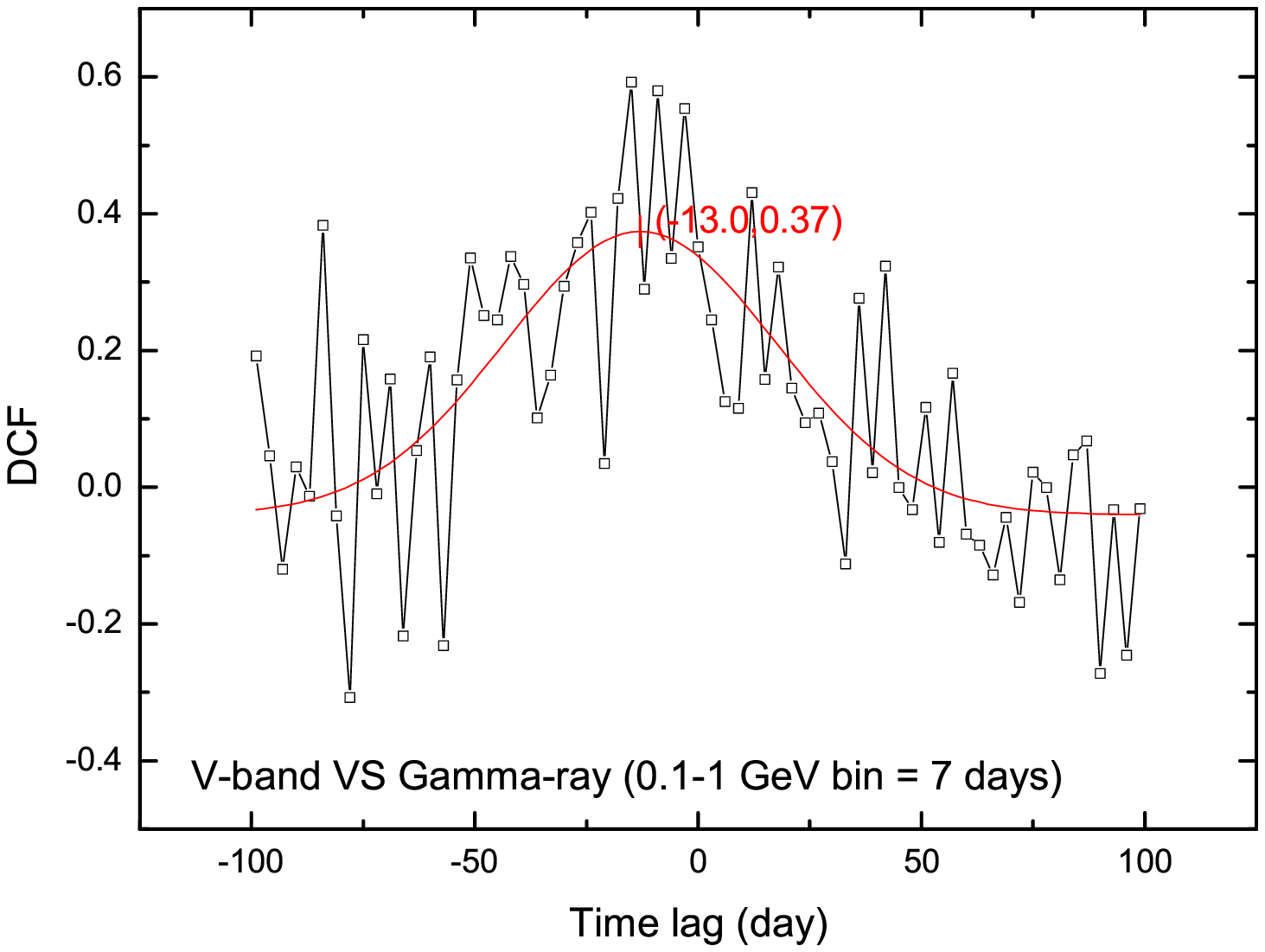}
\includegraphics[angle=0,scale=0.5]{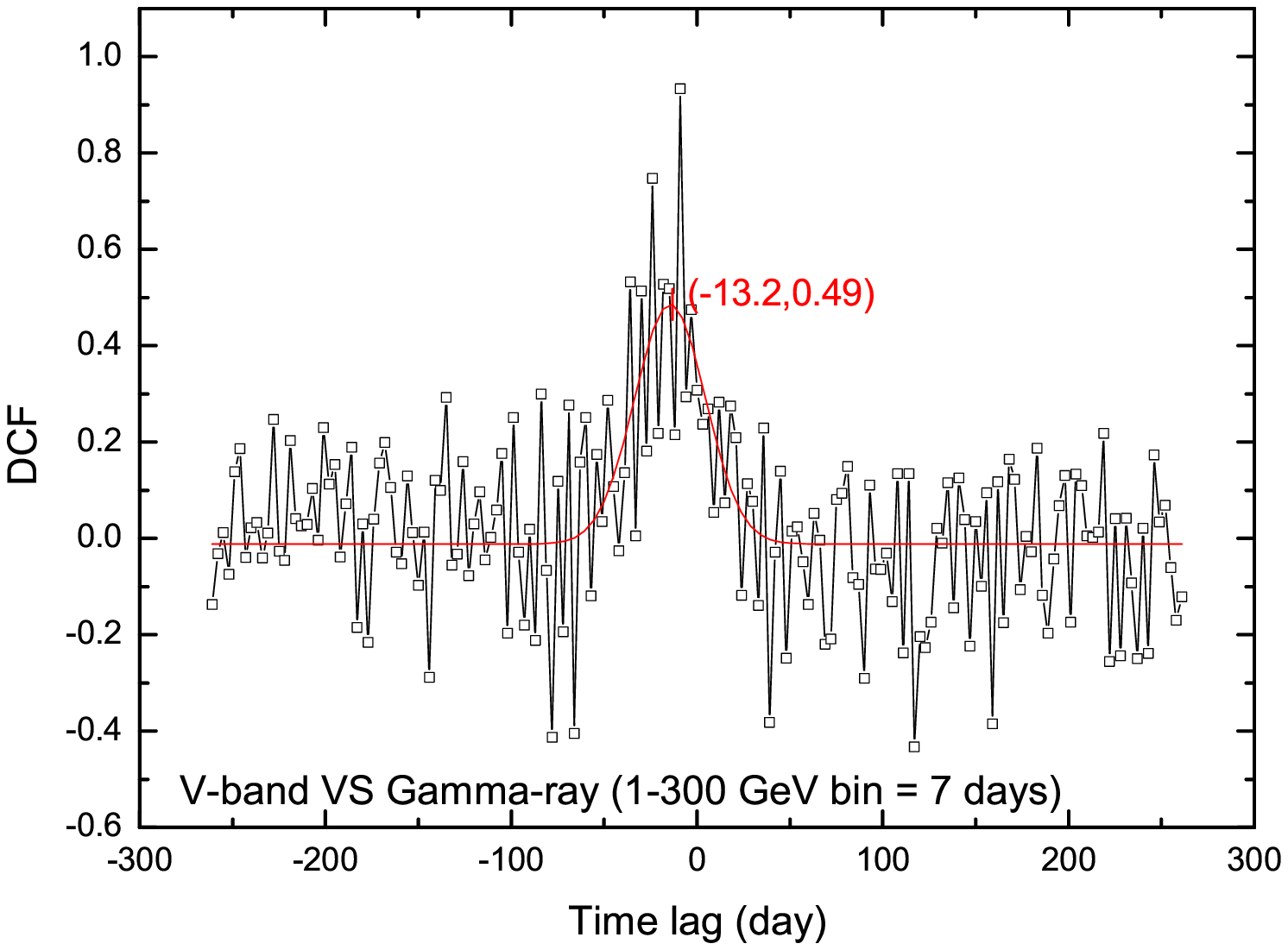}

\caption{The Discrete Correlation Function (DCF) results of $\gamma$-ray band with 7 days bin for 3FGL J0449.4-4350. }
\end{figure}

In order to investigate the possible QPO of this source, we use
the traditional Jurkevich (Jurkevich 1971) method to analyze the periodicity
information of the multi-wavelength band data. The Jurkevich (abbreviated as J-K) method
is based on the expected mean square deviation. It tests a run
of trial periods around which the data are folded, split into $m$ terms.
The $V^{2}_{m}$ reached its minimum means the trial period equal to a true one.
After analyzing the results of the statistical F-test, we get
a good guide to the fractional reduction of the variance, $f=\frac{1-V^{2}_{m}}{V^{2}_{m}}$, as in
Kidger et al. (1992). The higher the $f$ value, the higher the confidence of the period.
$f < 0.25$ usually indicates that the periodicity is a weak one with low confidence level.
The error in the period is estimated by
calculating the HWHM of the shape of
minimum in the $V^{2}_{m}$ curve, as shown in Fig. 4.
The periods indicated by the Jurkevich (J-K) method are also summarized in Table 2.
We also adopted the Discrete Auto-Correlation Function (DACF) method to the
multi-wavelength data set for possible QPO, as shown in Fig. 4. The DACF method is described in detail in Edelson \& Krolik (1988) and Hufnagel \& Bregman (1992) (also see our previous paper, Xie et al. 2008, Li, 2015). The periods indicated by the DACF method and the Coefficient (Coef.) are also listed in Table 2. The QPO of 450 days appears to have occurred, but it is very weak.
%From Table 2, we can find  $T3=1370$ ($\approx 3\times T1$) days and $T4=1215$ ($\approx 2\times T2$) days in DACF method.
%The $T1=450$ and $T2= 630$ days periods for optical and $\rm\gamma$-ray variability data set were confirmed by three kinds of period analysis methods. So we consider that the $T1$ and $T2$ is two possible quasi-periodic timescale. The proportional relationship of $T3:T1\approx3:1$ and $T4:T2\approx2:1$ also support this result, because an identical red noise frequency (or periodicity) almost cannot coincidentally appear in two light curve data (Vaughan 2005).

\section{CROSS-CORRELATION ANALYSIS}
As shown in Fig. 1, it seems that the $\gamma$-ray and optical fluxes (V band fluxes shown as magnitude) may
well have risen and declined together. Although the large time gaps in the optical V band light curve may limit
the interpretation, the correlations are obvious. In order to investigate the relationship between the
variability of $\gamma$-rays light curve and that in the optical V band light curve, we calculated
the discrete correlation functions (DCFs; Edelson \& Krolik 1988; Hufnagel \& Bregman 1992)
between these different wavelength-bands. An obvious peak in the DCF means a strong correlation
between two wavelength-bands, with a higher DCF peak corresponding to a stronger correlation.
As shown in Fig. 5, a peak at a lag of $\tau_{\rm peak} = -16.5\pm21.5$ (the error is HWHM of the peak) days with correlation coefficient DCF$_{\rm peak}$ = 1.10 (in left panel), and another peak at a lag of $\tau_{\rm peak}= -16.5\pm15.5$  days with correlation coefficient DCF$_{\rm peak}$ = 0.94 (in right panel) that indicates strong correlation between the optical variations and the $\gamma$-ray ones (Yoshida et al. 2019). Considering the time bin is 30 days, so the time lag is not too precise, with large error.
For checking the correlation and negative time lag, we analysis the relationship between the $\gamma$-ray (7 days time bin) and optical V-band, and also found the strong correlation with DCF$_{\rm peak}$ = 0.93 (as shown in upper right panel of Fig. 6) and negative lag with a Gauss fit mean value $\tau_{fit}=-13.1$ days (as shown in lower panels of Fig. 6). This indicate that the strong correlation and negative time lag are reliable.
The negative time lag indicate that the $\gamma$-ray activity (variability) slightly preceded the one of optical V band with a few days, and the emissions from differen bands originate from the same region (the same electron group). If the $\gamma$-ray emission is due to inverse-Compton scattering of soft photons by the same electrons producing synchrotron radiation (the radio, optical,and X-ray radiation), the $\gamma$-ray emission variations are expected to be simultaneous or delayed with respect to those characterising the optical radiation, as resulting from modelling non-thermal flares with shocks in a jet (e.g. Sikora et al. 2001; Sokolov et al. 2004; Sokolov \& Marscher 2005; Carnerero et al. 2015). However, if the soft photons is the far ultraviolet, the $\gamma$-ray emission activity may preceded the one of optical band, because the far ultraviolet radiation preceded the optical emission in the synchrotron radiation process.

\section{SUMMARY AND DISCUSSION}
3FGL J0449.4-4350 shows strong variations at both bands ($\gamma$-ray and optical V band), and these variations are highly correlated, with features in one band generally showing up in the other band as well. This is clearly illustrated in Fig. 1. This phenomenon may be attributed to the jet beaming effect, i.e. Doppler factor change with time. We calculated the DCFs of optical and $\gamma$-ray light curves, and found the correlation coefficient is very high, as shown in the Fig. 5 and Fig. 6. The time lag of optical V band to $\gamma$-ray can be explained by lepton synchrotron-self Compton (LSSC) model. However, there are two tentative radiation models to produce the $\gamma$-ray flux, i.e. lepton SSC model (Zhou et al. 2014) and proton synchrotron model (Psyn, Zhang, et al. 2012, 2013, 2018; Gao et al. 2018) for 3FGL J0449.4-4350. The strong correlations of the two different wavelength band of this source support the LSSC model. Further observations are needed to confirm the reliability of this conclusion.

We analyzed light curves of the $\gamma$-ray, optical V band of this source, using four methods, to show any indecations of quasi-periodicity, and found a possible QPO of $T1=450\pm23$ days in the light curves, but the confidence level is marginal. We devote our efforts in demonstrating the authenticity of this quasi-cycle, but it is difficult to distinguish periodic and stochastic signals in the light curve of blazar, because of the  steep spectrum ('red noise') stochastic processes can display few-cycle periodicity (Press 1978; Vaughan et al. 2016). As show in Fig. 1, a visual inspection indicated six cycles persistent throughout the light curve. However, the longer duration of the data sample is needed to examined the possibe QPO (Kidger et al. 1992; Vaughan et al. 2016; Kova{\v{c}}evi{\'c} et al. 2018). 3FGL J0449.4-4350 have approximate quasi-periodicity in $\gamma$-ray and optical bands, which is interesting. If the QPO of $~450$ days is real, which theoretical model can provide a reasonable explanation for it? So far, the several physical reasons often mentioned in the literature are as follows (Ackermann et al. 2015A, Zhou et al. 2018). (1)The observed year-like timescale periodicity of blazar could be related to the orbital motion of a central binary black hole system (Lehto \& Valtonen 1996; Yu 2001; Sandrinelli et al. 2014, 2016A; Zhang et al. 2017A, Graham et al. 2015A, B). In this scenario, the enough large secondary black hole is in an orbit which is non-coplanar with the accretion disk of a primary black hole, inducing torques in the inner parts of the disk and resulting in the precession of the disk and its jet.
(2)The supermassive binary black hole (SMBBH) system with thick disk excites the p-mode oscillations of the thick disk, subsequently the quasi-periodic injection of plasma from an oscillating accretion disk pour into the jet, therefore the jet emission produce the quasi-periodic flux (Liu et al. 2006, Liu \& Chen 2007).
(3)If the angular momentum of the inner disc is not paralleled to spin of the central supermassive black hole(SMBH), it would undergo Lense-Thirring precession around the SMBH spin (wilkins 1972; Gupta ea al. 2019). The direction of the associated jet also have precessional motion (Caproni \& Abraham 2004, Caproni et al. 2004, Li et al. 2010, Kudryavtseva et al. 2011), therefore produce flux vairation in each electromagnetic band (Wilkins 1972; Ackermann et al. 2015A).
(4)If the jet have helical large scale magnetic field, or itself is helical by hydrodynamical instabilities or by relativistic shocks (e.g, Marscher \& Gear 1985; Rieger 2004; Rani et al. 2009; Larionov et al. 2013; Raiteri et al. 2017; Hong et al. 2018) or by the turbulent in plasma of jet (Marscher 2014), the emission of the jet is also quasi-periodicity (Camenzind \& Krockenberger 1992, Villata \& Raiteri 1999; Gupta et al. 2019).

Considering that our source is TeV BL Lacs, the possible physical reason of the quasi-periodicity of $~450$ days is more likely to the helical motion of jet. Rieger (2004) found that the relationship between the observed quasi-periodicity $P$ and the real physical driving period $P_{d}$ of helical motion can be estimated by the formula,
\begin{equation}
P_{d} \simeq \frac{\gamma_{b}^{2}}{1+z}P,
\end{equation}
where $z$ is the redshift and $\gamma _{b}$ is the bulk Lorentz factor. For 3FGL J0449.4-4350, the bulk Lorentz factor is $\gamma _{b}\sim7.4$ ( Nemmen et al. 2012). Based on equation (4) and the possible quasi-periodicity $P = T_{1}=450$ days, the physical driving quasi-periodicty of helical motion of jet in 3FGL J0449.4-4350 is $P_{d}\simeq 55.99$ years. On the other hand, If $\gamma _{b}\sim15$ ( Henri \& Saug\'{e} 2006) was adopted, $P_{d}\simeq 230.05$ years.
The helical motion of the jet is most likely driven by the orbital motion in the SMBBH system, which imply that 3FGL J0449.4-4350 is a possible candidate of SMBBH.
According to cold dark matter (CDM) cosmology model, SMBBH is formed after the merger of galaxies. If the mass ratio between the primary and secondary black holes less than 3 , i.e. $R \leq\frac{3}{1} $, we call it "major merger"; if $\frac{3}{1}\leq R \leq 10^4$, it is "minor merger" (Kauffmann \& Haelnelt 2000; Springel et al. 2005).
For any given value of the mass ratio of SMBBH, the mass of the primary black hole can be estimated by the formula(Begelman et al. 1980; Ostorero et al. 2004; Li et al. 2015):
\begin{equation}
M\simeq P_{d}^{\frac{8}{5}} R^{\frac{3}{5}}\ \rm{M_{\bigodot}},
\end{equation}
where $P_{d}$ is the value of the QPO in unit of year.
For major merger of the SMBBH system, the mass ratio can be assumed to be $R=\frac{3}{2} $.
Based on our results $P_{d} = 55.99$ years, the mass of the primary black hole of 3FGL J0449.4-4350 is about $M\simeq 8.0 \times 10^{8} \ \rm{M_{\bigodot}}$. At the same time, If $\gamma _{b}\sim15$ ( Henri \& Saug\'{e} 2006) was adopted, $M\simeq 7.7 \times 10^{9} \ \rm{M_{\bigodot}}$
For minor merger of the SMBBH system, assuming $R=10$, if $\gamma _{b}\sim7.4$ and $\gamma _{b}\sim15$, the corresponding mass of the primary is
$M\simeq 2.5 \times 10^{9} \ \rm{M_{\bigodot}}$ and $M\simeq 2.4 \times 10^{10} \ \rm{M_{\bigodot}}$, respectively.

In the binary black hole theoretic model, the QPO can come from the perturbation of the accretion disk by the secondary black hole.
In this binary black hole system, assuming the QPO come from the rotating accretion disc, the mass of the primary black hole can be estimated by the formula (Abramowicz at al. 2004):
\begin{equation}
M\simeq 3\times 10^3 P\ \rm{M_{\bigodot}},
\end{equation}
where $P$ is the value of the QPO in unit of second. We generalize this formula to SMBBH.
For $P= T_{1}=450 \rm{\ days} = 3.888\times 10^8 \rm{\ seconds}$, $M=1.7\times 10^{11}\ \rm{M_{\bigodot}}$. This value of the mass of center black hole is too large. So the assumption that the QPO come from the rotating accretion disc may be inappropriate for this object.
In comparison, assuming the QPO come from the helical motion of jet is more preferable than assuming the QPO come from the rotating accretion disc. The black hole mass $M\simeq 7.7 \times 10^{9} \ \rm{M_{\bigodot}}$ based on bulk Lorentz factor $\gamma _{b}\sim15$ of the jet is more reasonable.

\section*{Acknowledgements}
We thank the anonymous referee for the valuable suggestions. This work is supported by the National Natural Science Foundation of China (grants 11863007, 11573034, 11851304), the joint fund of Astronomy of the National Nature Science Foundation of China and the Chinese Academy of Science (grant U1731239).The CSS survey is funded by the National Aeronautics and Space
Administration under Grant No.NNG05GF22G issued through the Science
Mission Directorate Near-Earth Objects Observations Program.
The CRTS survey is supported by the U.S. National Science Foundation under grants AST-0909182.
The authors are grateful for the above mentioned database and facilities.

%The generated time series have properties (length, sampling interval) similar to these data sets.
%A significant clustering of samples along the time axis may decrease the number of statistically independent frequencies to 1/3 N (Horne and Baliunas, 1986)

\end{document}